\documentclass[fleqn,usenatbib]{mnras}

\usepackage{newtxtext,newtxmath}

\usepackage[T1]{fontenc}

\newcommand{\nickel}{$\rm ^{56}Ni$}
\newcommand{\cobalt}{$\rm ^{56}Co$}

\newcommand{\msun}{$M_\odot$}
\newcommand{\rsun}{$R_\odot$}

\DeclareRobustCommand{\VAN}[3]{#2}
\let\VANthebibliography\thebibliography
\def\thebibliography{\DeclareRobustCommand{\VAN}[3]{##3}\VANthebibliography}

\usepackage{graphicx}	
\usepackage{amsmath}	

\title[SN\,2020fqv progenitor and CSM]{Progenitor and Close-In Circumstellar Medium of Type II Supernova 2020fqv from High-Cadence Photometry and Ultra-Rapid UV Spectroscopy}

\author[S. Tinyanont et al.]{Samaporn~Tinyanont,$^{1}$\thanks{E-mail: stinyanont@ucsc.edu}
R.~Ridden-Harper,$^{2}$
R.~J.~Foley,$^{1}$
V.~Morozova,$^{3}$
C.~D.~Kilpatrick,$^{4,5}$
G.~Dimitriadis,$^{1,6}$
\newauthor
L.~DeMarchi,$^{4,5}$
A.~Gagliano,$^{7,8}$
W.~V.~Jacobson-Gal\'an,$^{4,5}$
A.~Messick,$^{9}$
J.~D.~R.~Pierel,$^{10}$
A.~L.~Piro,$^{11}$
\newauthor
E.~Ramirez-Ruiz,$^{1}$
M.~R.~Siebert,$^{1}$
K.~C.~Chambers,$^{12}$
K.~E.~Clever,$^{1}$
D.~A.~Coulter,$^{1}$
K.~De,$^{13}$
\newauthor
M.~Hankins,$^{14}$
T.~Hung,$^{1}$
S.~W.~Jha,$^{15}$
C.~E.~Jimenez~Angel,$^{16,17}$ 
D.~O.~Jones,$^{1}$
M.~M.~Kasliwal,$^{13}$
C.-C.~Lin,$^{12}$
\newauthor
R.~Marques-Chaves,$^{18}$
R.~Margutti,$^{4,5}$
A.~Moore,$^{19}$
I.~P\'{e}rez-Fournon,$^{16,17}$
F.~Poidevin,$^{16,17}$
A.~Rest,$^{2,10}$
\newauthor
R.~Shirley,$^{20}$
C.~S.~Smith,$^{1}$
E.~Strasburger,$^{21}$
J.~J.~Swift,$^{22}$
R.~J.~Wainscoat,$^{12}$
Q.~Wang,$^{2}$
Y.~Zenati$^{2,23}$\\
\\
$^{1}$Department of Astronomy and Astrophysics, University of California, Santa Cruz, CA 95064, USA\\
$^{2}$Department of Physics and Astronomy, The Johns Hopkins University, Baltimore, MD 21218, USA \\
$^{3}$Department of Physics, The Pennsylvania State University, University Park, PA 16802-6300, USA \\
$^{4}$Center for Interdisciplinary Exploration and Research in Astrophysics (CIERA) and Department of Physics and Astronomy, \\ Northwestern University, Evanston, IL 60208, USA \\
$^{5}$Department of Physics and Astronomy, Northwestern University, 2145 Sheridan Road, Evanston, IL 60208, USA \\
$^{6}$School of Physics, Trinity College Dublin, The University of Dublin, Dublin 2, Ireland \\
$^{7}$Department of Astronomy, University of Illinois at Urbana-Champaign, 1002 W. Green St., IL 61801, USA \\
$^{8}$Center for Astrophysical Surveys, National Center for Supercomputing Applications, Urbana, IL, 61801, USA\\
$^{9}$Department of Physics \& Astronomy, Washington State University, Pullman, WA 99164, USA\\
$^{10}$Space Telescope Science Institute, 3700 San Martin Drive, Baltimore, MD 21218, USA \\
$^{11}$The Observatories of the Carnegie Institution for Science, 813 Santa Barbara St., Pasadena, CA 91101, USA \\
$^{12}$Institute for Astronomy, University of Hawaii, 2680 Woodlawn Drive, Honolulu, HI 96822, USA\\
$^{13}$Division of Physics, Mathematics and Astronomy, California Institute of Technology, Pasadena, CA 91125, USA\\
$^{14}$Arkansas Tech University, Russellville, AR 72801, USA\\
$^{15}$Department of Physics and Astronomy, Rutgers the State University of New Jersey, 136 Frelinghuysen Road, Piscataway, NJ 08854, USA \\
$^{16}$ Instituto de Astrof\'\i sica de Canarias, C/V\'\i a L\'actea, s/n, E-38205 San Crist\'obal de La Laguna, Tenerife, Spain\\
$^{17}$Universidad de La Laguna, Dpto. Astrof\'\i sica, E-38206 San Crist\'obal de La Laguna, Tenerife, Spain\\
$^{18}$Geneva Observatory, University of Geneva, Chemin Pegasi 51, CH-1290 Versoix, Switzerland \\
$^{19}$Research School of Astronomy and Astrophysics, Australian National University, Canberra, ACT 2611, Australia\\
$^{20}$Department of Physics and Astronomy, University of Southampton, Highfield, Southampton SO17 1BJ, UK \\
$^{21}$Department of Astronomy, University of California, Berkeley, CA 94720-3411, USA\\
$^{22}$Thacher Observatory, Thacher School, 5025 Thacher Rd. Ojai, CA 93023 USA\\
$^{23}$CHE Israel Excellence Fellowship
}
\date{Accepted XXX. Received YYY; in original form ZZZ}

\pubyear{2021}

\begin{document}
\label{firstpage}
\pagerange{\pageref{firstpage}--\pageref{lastpage}}
\maketitle

\begin{abstract}
We present observations of SN 2020fqv, a Virgo-cluster Type II core-collapse supernova (CCSN) with a high temporal resolution light curve from the \textit{Transiting Exoplanet Survey Satellite} (\textit{TESS}) covering the time of explosion; ultraviolet (UV) spectroscopy from the \textit{Hubble Space Telescope} (\textit{HST}) starting 3.3~days post-explosion; ground-based spectroscopic observations starting 1.1~days post-explosion; along with extensive photometric observations. 
Massive stars have complicated mass-loss histories leading up to their death as CCSNe, creating circumstellar medium (CSM) with which the SNe interact. Observations during the first few days post-explosion can provide important information about the mass-loss rate during the late stages of stellar evolution. Model fits to the quasi-bolometric light curve of SN 2020fqv reveal ~0.23 \msun\ of CSM confined within ~1450 \rsun\ ($10^{14}$ cm) from its progenitor star. Early spectra ($<$4 days post-explosion), both from \textit{HST} and ground-based observatories, show emission features from high-ionization metal species from the outer, optically thin part of this CSM. 
We find that the CSM is consistent with an eruption caused by the injection of $\sim${}$5\times 10^{46}$ erg into the stellar envelope $\sim$300 days pre-explosion, potentially from a nuclear burning instability at the onset of oxygen burning.
Light-curve fitting, nebular spectroscopy, and pre-explosion \textit{HST} imaging consistently point to a red supergiant (RSG) progenitor with $M_{\rm ZAMS}${}$\approx${}$13.5$--$15 \, M_{\odot}$, typical for SN~II progenitor stars. This finding demonstrates that a typical RSG, like the progenitor of SN\,2020fqv, has a complicated mass-loss history immediately before core collapse.
\end{abstract}




\section{Introduction}
Massive stars ($\gtrsim${}$8 \, M_\odot$) shed a significant amount of mass towards the end of their lives, forming circumstellar medium (CSM) with variable density profiles, physical extents, and total mass.
Physical processes responsible for mass loss may include stellar winds, minor eruptive mass loss associated with late-stage nuclear burning instabilities, binary interactions, and likely combinations thereof \citep[][and references therein]{smith2014}.
However, the rates and quantitative contributions from each of these processes in different types of progenitor stars remain the subject of ongoing research (see, e.g., pre-SN instability and outburst, \citealp{wu2021, leung2020}; binary effects on stellar structure, \citealp{laplace2020,zapartas2021}; and new stellar wind prescriptions \citealp{bjorklund2021, kee2021}). 
Observations of the resulting core-collapse supernova (CCSN) interacting with the CSM can probe its density structure, providing clues about its origin and the properties of the progenitor star. 

CCSNe with a large amount of CSM (a few $M_\odot$) close to the progenitor star have been observed for decades, as they are luminous and show persistent interaction signatures \citep{schlegel1990}. 
In this scenario, the SN shock collides with the CSM and converts kinetic energy into heat, which gets radiated away as extra luminosity. 
The spectra of these SNe are classified as Type IIn \citep{schlegel1990, filippenko1997} with strong and persistent (over months or years) narrow ($\sim$100 -- 1000 ~$\rm km\,s^{-1}$) recombination lines from hydrogen in the CSM.
There is a rarer class of strongly interacting SNe with only helium lines called Ibn \citep{matheson2000, pastorello2007, foley2007, pastorello2008}, with a smaller associated CSM mass. 
More recently, events with a CSM lacking hydrogen and helium are found and classified as Type Icn \citep{gal-yam2021}. 
In addition, some stripped-envelope SNe have been observed to exhibit Type IIn-like spectra at late times, several months post-explosion, indicative of mass loss centuries before the explosion (e.g., SNe\,2001em, \citealp{chugai2006}, \citealp{chandra2020}; 2004dk, \citealp{mauerhan2018}; 2014C, \citealp{milisavljevic2015, margutti2017}; 2019oys, \citealp{sollerman2020}, and 2019yvr, Auchettl et al. in prep).
These strongly interacting SNe are also bright and long-lasting in the infrared (IR), with some remaining detected decades after the explosion (e.g., \citealp{fox2011, tinyanont2016,  tinyanont2019c}). 
While readily detectable, these strongly interacting SNe represent only about 10\% of all CCSNe \citep{smith2011}.
They are products of progenitor systems with the most extreme mass loss, such as Luminous Blue Variables (LBVs), extreme red supergiants (RSGs) with eruptive mass loss, or interacting binary systems \citep[][and references therein]{smith2009,foley2011, margutti2014, smith2017}. 

For the majority of CCSNe, weaker signs of CSM interaction are present but have escaped scrutiny for decades until the recent advent of large-scale transient surveys that discover a large number of SNe, some at very early times. 
These surveys include the All-Sky Automated Survey for SuperNovae (ASAS-SN; \citealp{shappee2014}), the Asteroid Terrestrial-impact Last Alert System (ATLAS; \citealp{tonry2018}), the Young Supernova Experiment (YSE; \citealp{jones2021}), and the Zwicky Transient Facility (ZTF; \citealp{bellm2019}).
For dense and nearby CSM from mass loss immediately before the SN, the energetic shock breakout (SBO) emission radiatively ionizes the CSM, producing narrow Balmer series, \ion{He}{i}, and highly-ionized metallic emission lines with pronounced electron-scattering Lorentzian wings.
These so-called flash ionization features only last for hours to days as the CSM recombines or gets overrun by the SN shock; thus, they are missed in SNe with no early observations.
Examples of CCSNe exhibiting these features include SNe\,2013cu \citep{galyam2014}, 2013fs \citep{yaron2017}, 2017ahn \citep{tartaglia2020}, 2020pni \citep{terreran2021}, and 2020tlf (Jacobson-Gal\'{a}n et al., in prep).

While the flash ionization features are ephemeral, CSM interaction can produce excess flux in the light curve of the SN up to $\sim$month post-explosion.
Such early-time bolometric luminosity excess has been observed in many CCSNe, including the nearby and well-studied SNe\,2017eaw \citep{morozova2020} and 2018cuf \citep{dong2020}.
In some events, like SN\,2017gmr \citep{andrews2019}, the excess flux from the CSM can be present without early-time flash ionization features.
\cite{morozova2018} demonstrated using light curves of 20 nearby hydrogen-rich (Type II) SNe that early-time excess flux from CSM interaction is a generic feature of this class of CCSNe.
In addition to photometric evidence, there is also spectroscopic evidence of high-velocity absorption features from hydrogen and helium caused by the continuous excitation of the outermost ejecta by ongoing CSM interactions \citep{chugai2007} in many SNe II-P (SNe II with a luminosity plateau) \citep[e.g.,][]{gutirrez2017, tinyanont2019b, davis2019, dastidar2019, dong2020}.
Because SNe II-P are explosions of normal RSGs (see review by \citealp{smartt2009}), the ubiquity of CSM interaction in SNe II-P indicates enhanced mass loss in these stars towards the end of their life, perhaps due to the instability in late-stage nuclear burning \citep{quataert2012, fuller2017}.

Observing and constraining CSM properties of SNe II-P can help us better understand the evolution of RSGs prior to their death, and help us answer some unresolved problems in massive stellar evolution. 
For instance, one of the most contentious issues is the ``RSG problem'', which is the apparent lack of high-mass ($> 17$ \msun) RSG progenitors to SNe II-P, despite the presence of RSGs in this mass range in the local universe \citep{smartt2009IIP, Smartt15}.
There are many possible solutions to this problem.
RSGs in this mass range may experience a direct collapse into a black hole, producing no SN \citep[e.g.,][]{oconnor2011, adams2017, sukhbold2020}.
There may be biases with progenitor mass measurements from pre-explosion imaging alone; some studies inferring the progenitor mass from star formation histories found that high-mass RSGs can produce CCSNe \citep[e.g.,][]{jennings2014, auchettl2019}.
Alternatively, RSGs with high $M_{\rm ZAMS}$ may shed their hydrogen envelopes and evolve to another stellar type before explosion, explode in environments with high circumstellar or interstellar extinction \citep{walmswell2012,kilpatrick2018}, explode as another CCSN type \citep[e.g., SN~IIn][]{smith2014}, or some combination of all of these.
Relatedly, the explosions of stars with high $M_{\rm ZAMS}$ may be veiled by the very same dusty CSM and are missed by optical transient surveys \citep{jencson2019}.
Lastly, \citet{davies2018, davies2020} argue that the RSG problem is not statistically significant, since there are considerable uncertainties in the mass (and luminosity) measurements of progenitor RSGs, especially in the sampling of their spectral energy distributions and the bolometric corrections assumed.
{To further complicate this discussion, binary interaction, which is common in massive stars including RSGs, affects the mass evolution of a SN progenitor from $M_{\rm ZAMS}$ to the pre-explosion mass \citep{zapartas2019, zapartas2021}. }
Thus, a continued effort to follow-up SNe II-P to constrain their progenitor properties, in conjunction with surveys to find disappearing RSGs \citep[e.g.,][]{adams2017}, is required to understand the fate of high-mass RSGs.

\begin{figure*}
    \centering
    \includegraphics[width = 0.7\linewidth]{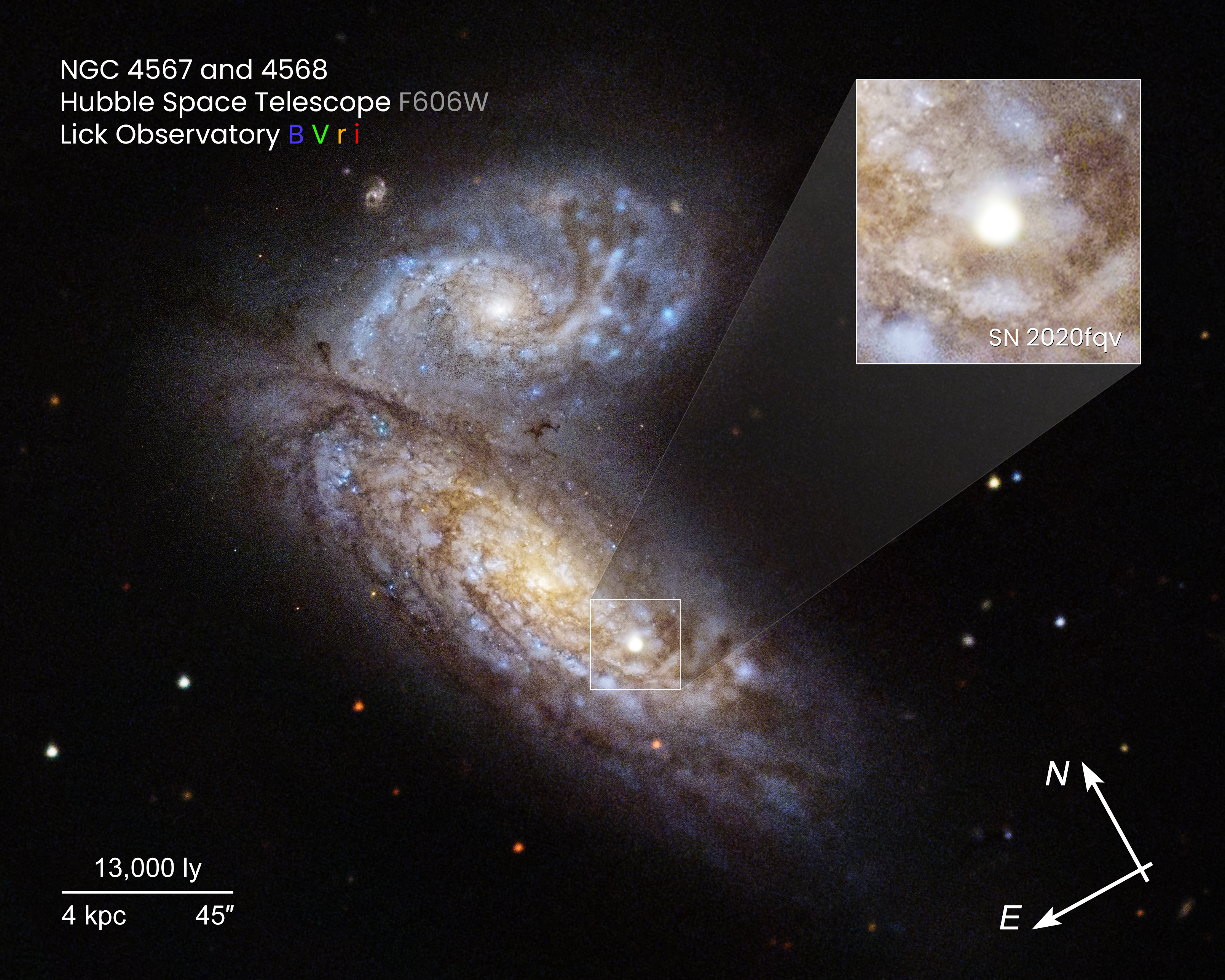}
    \caption{False-color image of SN\,2020fqv in its host galaxy system NGC\,4567 and 4568. The location of the SN is shown in the inset. The image is composed of a pre-explosion single-band \textit{HST}/WFPC2 image in the F606W filter wih the color information from ground-based images obtained using the Nickel Telescope at Lick Observatory. The image of the SN and the area outside of \textit{HST} field of view is also from Nickel. Image credit: Joseph Depasquale/STScI.}
    \label{fig:SN_color}
\end{figure*}

Here we present observations and data analysis of the nearby Type II-P SN\,2020fqv, focusing on its progenitor star and CSM properties. 
The SN was discovered on 2020 April 1 (ZTF20aatzhhl; \citealp{forster2020}; UT time used throughout the paper), in NGC\,4568.
The SN was monitored by the \textit{Transiting Exoplanet Survey Satellite} (\textit{TESS}; \citealp{ricker2015}), covering its explosion and subsequent rise. 
While it was originally classified as a SN II \citep[likely IIb;][]{zhang2020}, our light curves (Sec.~\ref{sec:photometry}) show a distinct luminosity plateau and our spectra (Sec.~\ref{sec:optical_spec}) show hydrogen at all phases, indicating that it is a Type~II-P SN.
We executed ultra-rapid target of opportunity (ToO) observations of the SN with the \textit{Hubble Space Telescope} (\textit{HST}); the first ultra-rapid ToO observations ever performed. 
We note that all epochs mentioned in this paper are relative to the explosion date derived in Sec.~\ref{sec:exp_epoch}. 
Fig.~\ref{fig:SN_color} shows the location of the SN in the host galaxy. 

The distance to NGC\,4568 was not well constrained in the studies of previous CCSNe\,1990B (Ib) and 2004cc (Ic) \citep[e.g.,][]{vandyk1993, clocchiatti2001, vandenbergh2005}, with assumed distances ranging from the distance to the Virgo cluster ($\sim$16~Mpc) to the distance based on its redshift ($\sim$32~Mpc).  
For this work, we adopt a distance derived from the Tully-Fisher relation in the near-IR of $17.3\pm3.6$~Mpc \citep{theureau2007}, consistent with being a Virgo cluster member.\footnote{We note that while \cite{theureau2007} lists a corrected kinematical distance to NGC\,4568 as 32.2~Mpc, they commented that the method they used to derive this figure assumes the redshift distance as a starting point and does not work well for galaxies in clusters.}
Given SN\,2020fqv's proximity and early detection, this SN presents another opportunity to probe the properties of close-in CSM around a common SN II-P. 

In Sec.~\ref{sec:obs}, we summarize the observations obtained for this SN. 
{In Sec.~\ref{sec:analysis}, we present different data analyses performed on the SN data.
We discuss the construction of the quasi-bolometric light curve and the analysis of the \textit{TESS} light curve. 
We determine the explosion and CSM parameters from our photometric observations.
We analyse early-time UV spectra and optical nebular spectra. 
Finally, we provide discussions and conclusions in Sec.~\ref{sec:conclusion}.} 


\section{Observations} \label{sec:obs}

\subsection{Rapid Early-Time Observations}\label{sec:early_time}
\begin{figure*}
    \centering
    \includegraphics[width=0.8\linewidth]{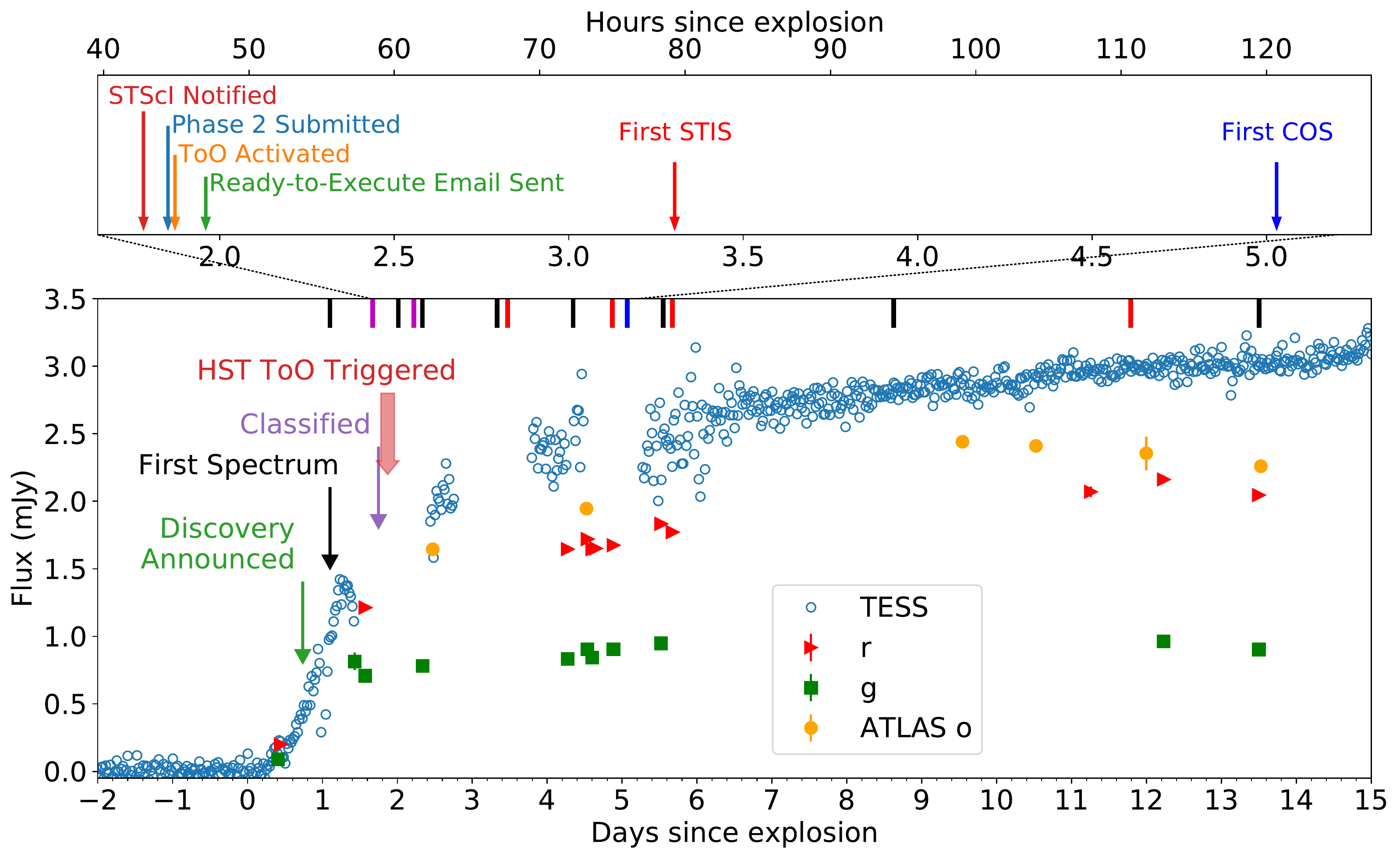}
    \caption{Early photometry of SN\,2020fqv from \textit{TESS}, along with select early ground-based photometric observations. 
    We mark the timeline of the events, including the discovery announcement at 2020 Mar 31, 15:59 (18 hr post-explosion); our first spectrum at 2020 Apr 01, 00:44 (26 hr post-explosion); classification on TNS at 2020 Apr 01, 16:16 (42 hr post-explosion; \protect\citealp{zhang2020}); and our \textit{HST} ToO trigger at 2020 Apr 01, 21:21, which leads to the first \textit{HST} STIS spectrum obtained at 2020 Apr 03, 05:36, 79 hours after the explosion. 
    Black ticks on the top of the bottom panel mark epochs of ground-based spectra in this work; red ticks are STIS spectra; and the blue tick is the one epoch of COS observations. 
    Magenta ticks mark the epochs of the public spectra available on TNS for this object.
    The width of the arrow marked ``HST ToO Triggered" in the main plot corresponds to the time span between ``STScI Notified" to ``Ready-to-Execute Email Sent'' in the top inset. }
    \label{fig:tess_photometry}
\end{figure*}

Immediately after the discovery and classification of SN\,2020fqv as a young nearby SN II in an active \textit{TESS} sector were announced, we activated our follow-up observation resources to capture the first few days of its evolution. 
We obtained the first optical spectrum on 2020 Apr 01 00:44 (26 hr post-explosion).
Following this and the public classification of SN\,2020fqv at 2020 Apr 01 16:15 \citep{zhang2020} (42 hr post-explosion), we notified the Space Telescope Science Institute (STScI) within 50 minutes (at 17:04 on the same day) of our intention to trigger our disruptive target of opportunity (ToO) program to observe it with \textit{HST}.
The phase 2 observation plans were submitted at 18:45, and the ToO trigger was submitted at 19:14. 
Finally, we informed STScI that the observations were ready to be executed at 21:21.
The first \textit{HST}/STIS observation began on 2020 Apr 03 at 05:36, a mere 32 hours after the ToO trigger was submitted, 37 hours after the SN's classification, and 79 hours after the explosion. 
\autoref{fig:tess_photometry} summarizes the timeline of the early observational sequence of SN\,2020fqv along with the light curves from \textit{TESS} and ground-based observatories described later in this Section.


\subsection{Ultraviolet Spectroscopy}\label{sec:uv_spec}
We triggered ultra-rapid UV spectroscopy observations of SN\,2020fqv with \textit{HST} as part of the Ultra-Rapid UV Spectroscopy of an Interacting Supernova Discovered by \textit{TESS} program (GO-15876; PI Foley) due to its proximity and youth at the time of discovery along with its location within \textit{TESS}'s active sector and signs of interaction from the classification spectrum \citep{dimitriadis2020}. 
\textit{HST} observed the SN at five epochs from 3 to 17 days post-explosion using the Space Telescope Imaging Spectrograph (STIS) and Cosmic Origin Spectrograph (COS), providing the wavelength coverage between 1300 and 4000 \AA. 
\autoref{tab:hst_log} summarizes the \textit{HST} observations. 
The COS observations were available only for day 5, in two different gratings. 
Another epoch of COS observation was attempted on day 11, but resulted in no data.

\begin{figure*}
    \centering
    \includegraphics[width=0.7\linewidth]{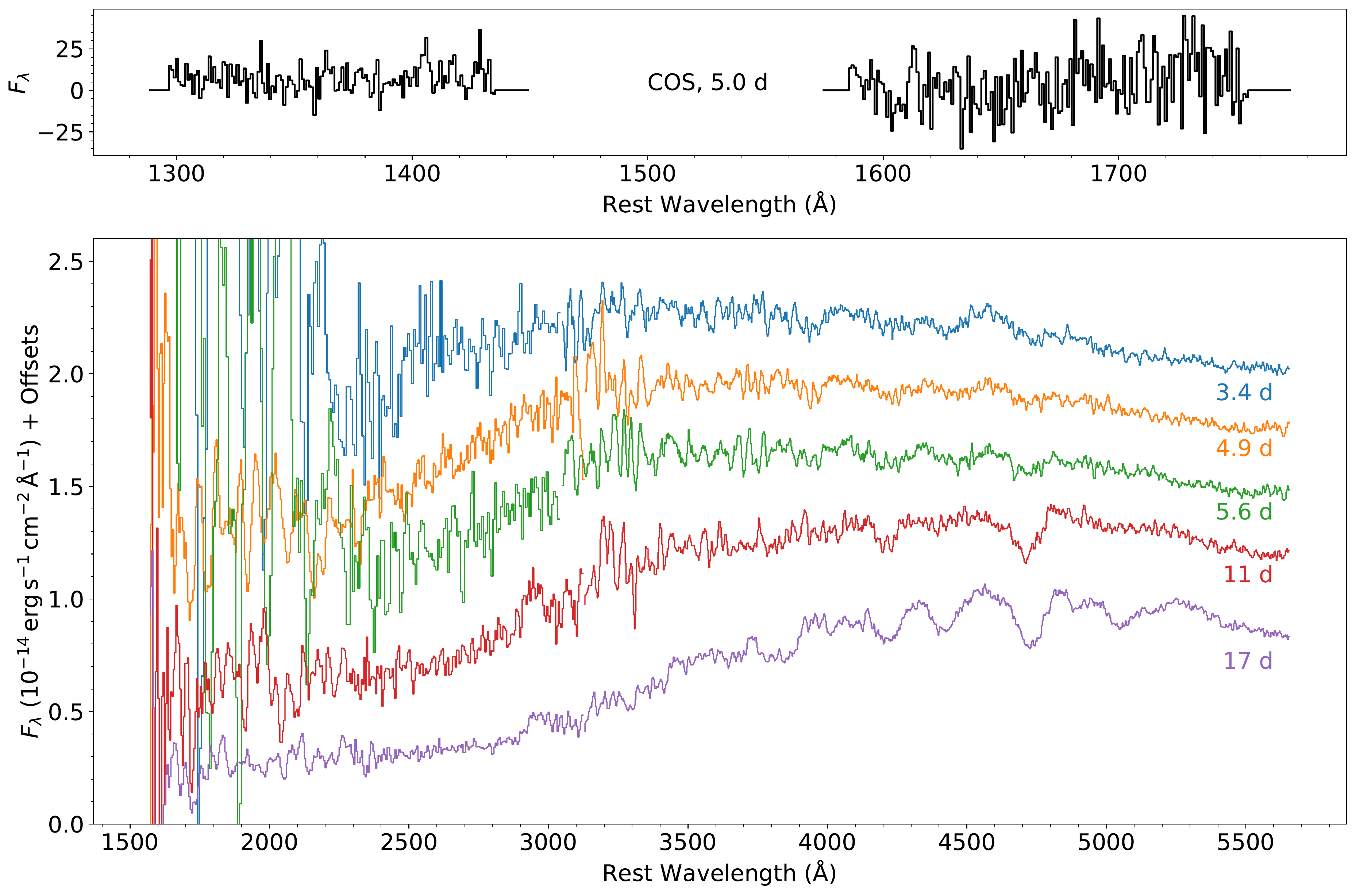}
    \caption{COS (top panel) and STIS (bottom panel) spectra of SN\,2020fqv from 3 to 17 days post-explosion. Fluxes in both panels are in the same units. COS spectra were taken at 5 days post-explosion with the G130M and G160M gratings in the FUV mode. There was no detection. For STIS, all the red channel data were taken using the CCD and the G430L grating. The blue channel data on days 3.4 and 5.6 were taken with the CCD and the G230LB grating, the rest were taken with the NUV-MAMA detector and the G230L grating.
    The spectra were corrected for dust extinction using the parameters derived in Sec.~\ref{sec:extinction}.   
    The red ($\lambda \gtrsim 3000 \,$ \AA) part of the spectra and the blue MAMA data were binned by 3 pixels and the blue CCD data were binned by 5 pixels to improve the S/N. 
    The uncorrected spectra are available via WISeREP. 
    }
    \label{fig:uv_spec}
\end{figure*}

The data were reduced using the standard STIS and COS data reduction pipelines, \texttt{stistools}\footnote{\url{https://stistools.readthedocs.io/}} and \texttt{calcos}\footnote{\url{https://github.com/spacetelescope/calcos}}, respectively. 
\autoref{fig:uv_spec} shows the STIS and COS spectra. 
For STIS, the blue channel spectra on days 3.4 and 5.6 were taken with the CCD and the G230LB grating while those on days 4.9, 11, and 17 were taken with the NUV-MAMA detector and the G230L grating. 
The spectra were corrected for dust extinction using the parameters derived in Sec.~\ref{sec:extinction} and the \cite{fitzpatrick1999} dust extinction law.
We used the package \texttt{extinction} to deredden the data\footnote{\url{https://extinction.readthedocs.io/}}. 
The spectra shown are binned for visualization.
The red ($\lambda \gtrsim 3000$ \AA) and the blue MAMA data were binned by 3 pixels while the blue CCD data were binned by 5 pixels.  
The COS spectra were obtained on day 5 using the G130M and G160M gratings. 
They only show non-detection.

\begin{table}
\caption{Log of \textit{HST} observations of SN\,2020fqv.}
\label{tab:hst_log}
\begin{tabular}{llllll}
\hline
Date       & MJD   & Epoch& Instrument/ & Grating & Exp. Time \\
           &       & (day)& Detector& & (s) \\
\hline
2020-04-03 & 58942.31 & 3.38     & STIS/CCD       & G230LB  & 6297          \\
2020-04-03 & 58942.45 & 3.52     & STIS/CCD       & G430L   & 920           \\
2020-04-04 & 58943.89 & 4.96     & STIS/NUV       & G230L   & 964           \\
2020-04-04 & 58943.74 & 4.81     & STIS/CCD       & G430L   & 1788          \\
2020-04-04 & 58943.96 & 5.03     & COS/FUV        & G160M   & 2032          \\
2020-04-05 & 58944.02 & 5.09     & COS/FUV        & G130M   & 5120          \\
2020-04-05 & 58944.54 & 5.61     & STIS/CCD      & G230LB  & 5089          \\
2020-04-05 & 58944.64 & 5.71     & STIS/CCD      & G430L   & 920           \\
2020-04-11 & 58950.71 & 11.8     & STIS/NUV        & G230L   & 964           \\
2020-04-11 & 58950.73 & 11.8     & STIS/CCD        & G430L   & 868           \\
2020-04-17 & 58956.69 & 17.8     & STIS/NUV        & G230L   & 5215          \\
2020-04-17 & 58956.81 & 17.9     & STIS/CCD       & G430L   & 1702  \\    
\hline
\end{tabular}
STIS/NUV refers to the NUV-MAMA detector
\end{table}









\subsection{\textit{TESS} Photometry}\label{sec:tess_photometry}
SN\,2020fqv occurred within camera 1 CCD 3 of \textit{TESS} during sector 23 which observed at 30 minute cadence from 58928 to 58954 MJD. The rise of SN\,2020fqv coincided with the end of orbit 53 and the start of orbit 54, during which time camera 1 suffered from extreme levels of scattered light from the Earth and Moon\footnote{\url{https://archive.stsci.edu/missions/tess/doc/tess_drn/tess_sector_23_drn32_v03.pdf}}. This extreme background led to detector saturation and data loss in camera 1 in the following two periods: 58941.68 to 58942.73~MJD and 58943.41 to 58944.20~MJD.

\begin{figure*}
    \centering
    \includegraphics[width=\textwidth]{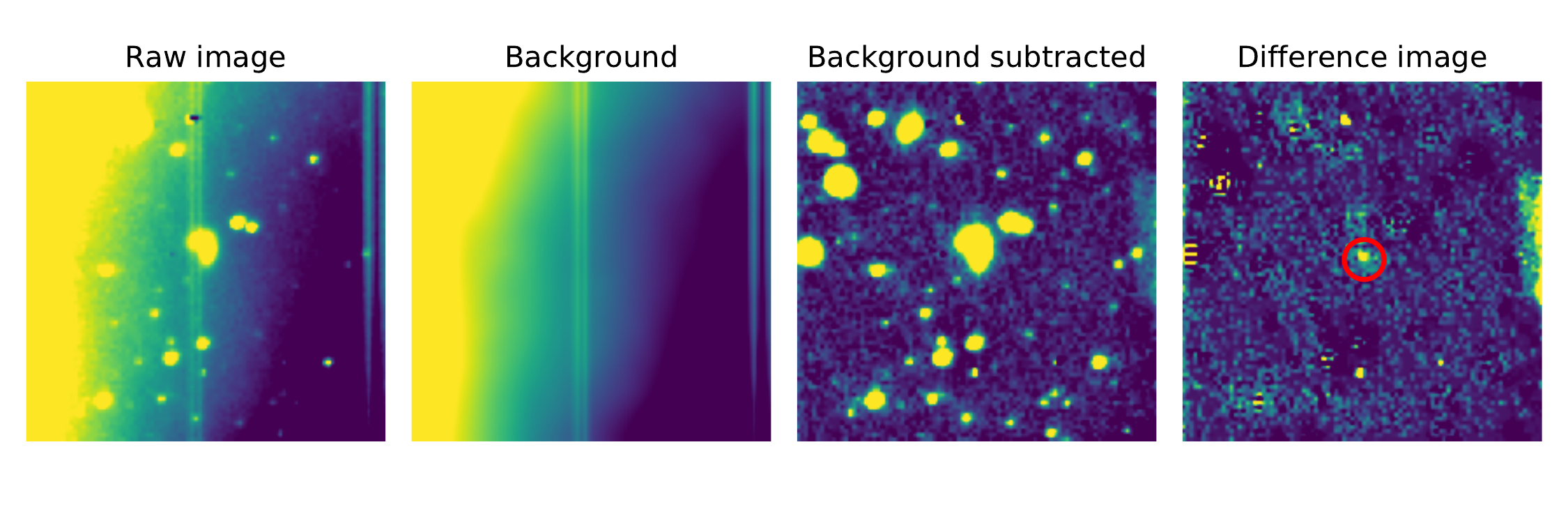}
    \vspace{-1cm}
    \caption{Reduction stages for a high background \textit{TESS} $90\times90$~pixel {($31\arcmin \times31\arcmin$)} image cutout centered on SN~2020fqv. The presence of extreme scattered light, detector straps and a spatially varying host requires a precise reduction method. 
    The leftmost panel shows the raw image from \textit{TESS}.
    The centre left panel shows the instrument background model including the scattered light and the detector strap.
    The centre right panel shows the sky image after background subtraction.
    The rightmost panel shows the difference image whereby a pre-explosion image is used to subtract the static host galaxy, revealing SN\,2020fqv in the red circle. 
    }
    \label{fig:tess_red}
\end{figure*}

Reducing \textit{TESS} data for SN\,2020fqv was further complicated by the source falling on columns that contain a feature of the detector known as a ``strap''. As described in the \textit{TESS} handbook\footnote{\url{https://archive.stsci.edu/files/live/sites/mast/files/home/missions-and-data/active-missions/tess/_documents/TESS_Instrument_Handbook_v0.1.pdf}} the straps scatter IR light that initially passes through the detector back into the detector, this has a colour dependent effect of enhancing the quantum efficiency of strap columns and neighbouring columns. This complex background feature, alongside the extreme levels of scattered light, which can be seen in \autoref{fig:tess_red}, at the rise of SN\,2020fqv presented a substantial challenge to extracting a clean \textit{TESS} light curve. 

To address these challenging data, we developed a data reduction pipeline for \textit{TESS} that reliably determines the background. This pipeline, known as \texttt{TESSreduce}, is publicly available on GitHub\footnote{\url{https://github.com/CheerfulUser/TESSreduce}}. While other \textit{TESS} difference imaging pipelines such as those presented in \citet{vallely2021, Bouma2019} and \citet{Woods2021} can produce a smoothed background, they fail to correctly account for the straps. Since SN\,2020fqv lies close to prominent straps we developed the following pipeline to correct for both the smooth background as well as the strap background.

To determine the complex \textit{TESS} background we first mask all known sources. We query the Gaia and PS1 source catalogues through CASjobs to a depth of 19$^{\rm th}$ magnitude and for each source scale the mask according to the \textit{G} band magnitude for Gaia, and \textit{i} band magnitude for PS1. The limit of 19$^{\rm th}$ magnitude is chosen as it is close to the \textit{TESS} detection limit, while still allowing enough pixels for background determination. We also mask all known transients in the image, such as SN\,2020fqv. Since the source catalogues do not contain accurate galaxy morphology information, we augment the source mask by adding remaining bright pixels to the mask that are identified by sigma clipping the source masked \textit{TESS} image for a low background image. This final mask is taken as the total source mask for all images. 

With all sources masked, we break the background determination into two components: a smooth continuous background and a discrete background defined by the detector straps. Alongside the source mask, we also construct a strap mask to mask out the strap columns and include the 3 neighbouring columns as a buffer zone. To construct the smooth background we mask out the strap regions and for each frame we interpolate the background over all masked pixels using the \texttt{scipy} \texttt{gridspec} with a linear method. We then smooth the background using a Gaussian kernel with a standard deviation of 3 pixels. 

The strap background is determined by identifying the effective quantum efficiency enhancement that the strap provides in each frame. We identify the enhancement for each column by masking all known sources and then calculate the median of the strap column counts divided by the smooth background which is interpolated over the strap columns. For small regions (e.g., $90\times90$~pixels), or times with little scattered light the scaling factor is constant for each column. To get the full background we multiply the strap scaling, as seen in \autoref{fig:tess_red}, with the smooth background. This background is largely free of biases from sources and artifacts like straps, allowing us to reliably subtract the extreme background present in sector 23 camera 1 of \textit{TESS} data.

Another crucial component to reduce \textit{TESS} data for SN\,2020fqv is to account for any pointing drift. Although \textit{TESS} has excellent temporal resolution, it has coarse spatial resolution with 21$''$ pixels. 
Even with \textit{TESS}'s relatively stable pointing, for spatially complex targets like NGC4568, small shifts of 0.01~pixels can lead to substantial changes to the counts contained in an aperture. For each image we identify the shift relative to a low background reference image using the \texttt{photutils} \texttt{DAOStarFinder} routine \citep{bradley2020}, the images are then aligned using the \texttt{scipy} \texttt{ndimage} \texttt{shift} routine \citep{2020SciPy-NMeth}. This alignment procedure reduces signals produced by telescope motion, allowing for effective difference imaging. 

Following the background subtraction and image alignment we calculate flux through standard aperture photometry. We use a $3\times3$~pixel source aperture and an annulus sky aperture, the resulting light curve is in un-calibrated \textit{TESS} counts. The calibration of \textit{TESS} counts to physical units is presented in Section~\ref{sec:tess_cal}.

\subsection{\textit{TESS} Photometric Calibration} \label{sec:tess_cal}
\textit{TESS} features a  broadband red filter that covers a wavelength range of 5802.57 to 11171.45 \AA. 
We calibrate the SN\,2020fqv \textit{TESS} light curve using synthetic photometry to the FLOYDS spectra taken during the rise which covers a wavelength range of 3200 to 10000 \AA. Since the FLOYDS spectra does not completely cover the \textit{TESS} bandpass, we extrapolate the spectra to 11500~\AA, using black body spectrum fit to each spectra. Furthermore, we smooth the original spectra using the \texttt{scipy} Savitzky-Golay filter, with a window length of 21 wavelength bins and a 3rd order polynomial. Although there is considerable uncertainty in the extrapolated region, the range we extrapolate over coincides with the decline to the red cutoff of the \textit{TESS} bandpass, reducing the overall impact.

We flux calibrate the FLOYDS spectra to coincidental \textit{o} band observations from ATLAS. We use FLOYDS spectra taken at 58941.26 MJD and 58943.28 MJD since they have ATLAS observations within 0.2~days, limiting the evolution that occurs between the data. We calculate the synthetic magnitude of the ATLAS o band for both spectra and normalise the spectrum such that the synthetic spectra equal the observed. Using the flux calibrated FLOYDS spectra we calculate the synthetic \textit{TESS} magnitude using the \textit{TESS} bandpass available on SVO \citep{SVO2012,SVO2020} and algorithms in the \texttt{pysynphot} package \citep{pysynphot}. The FLOYDS spectra used, alongside the \textit{TESS} and ATLAS o bandpasses are shown in \autoref{fig:tess_flux_cal}.

Finally the zeropoint is calculated by comparing the synthetic magnitudes, shown in Table~\ref{tab:tess_cal}, to the 6 hour averaged \textit{TESS} light curve. We find an AB zeropoint of $zp=20.81\pm 0.02$.
\autoref{fig:tess_photometry} shows the final \textit{TESS} photometry, along with our early observational timeline. 

\begin{table}
\caption{\label{tab:tess_cal} Observed ATLAS and synthetic \textit{TESS} magnitudes used for photometric calibration.}
\begin{tabular}{lll}
\hline
MJD   & ATLAS o (mag) & \textit{TESS} (syn-mag)  \\
\hline
58941.26    & 15.86 & 15.66 \\
58943.28    & 15.68 & 15.48 \\
\hline
\end{tabular}
\end{table}

\subsection{Ground-based Photometry}\label{sec:photometry}
\begin{figure*}
    \centering
    \includegraphics[width=0.7\linewidth]{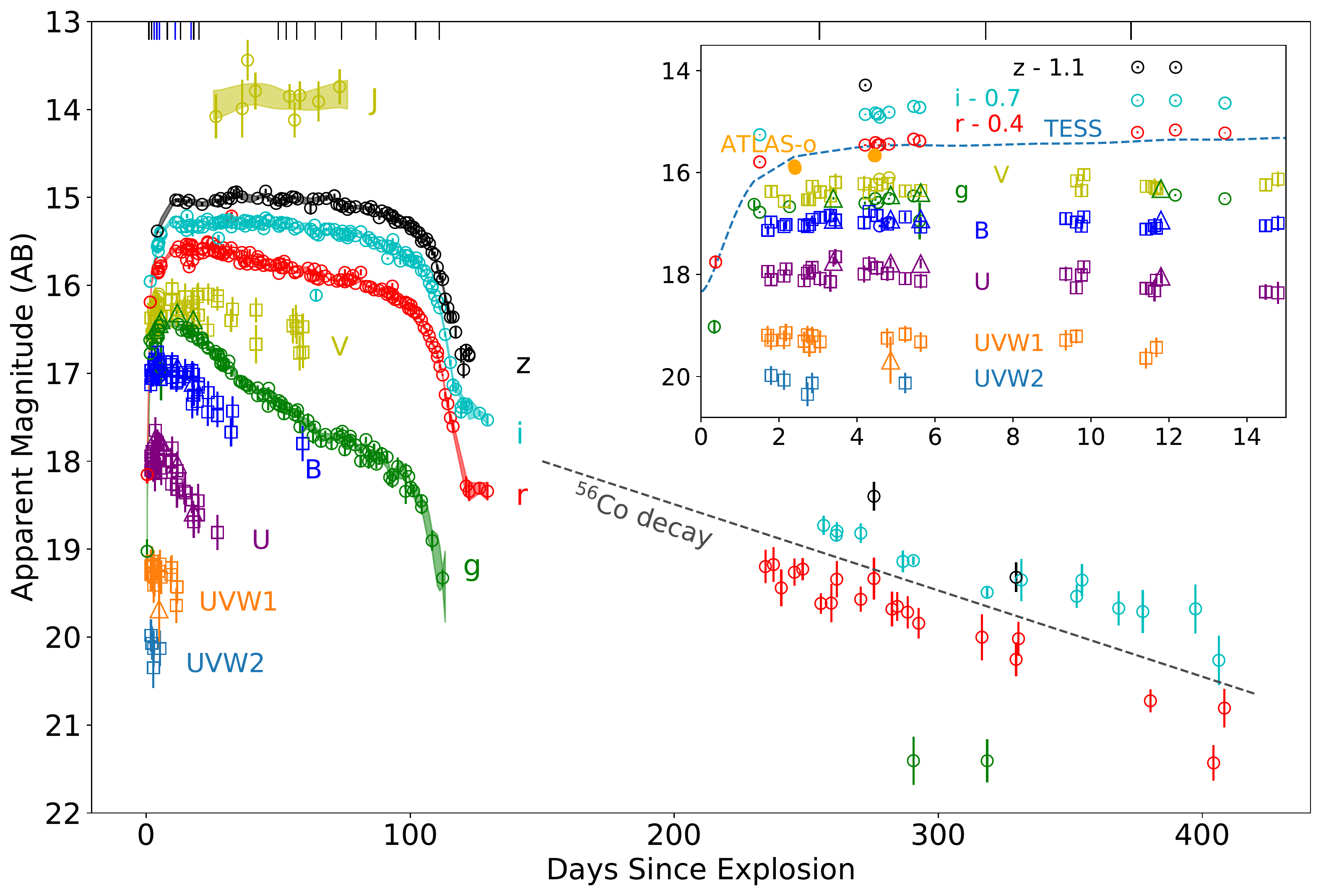}
    \caption{Photometry of SN\,2020fqv in the \textit{UVW1}, \textit{UVW2}, and \textit{UBVgrizJ} bands. 
    Circles are ground-based photometry; squares are photometry from \textit{Swift}; triangles are synthetic photometry from \textit{HST}/STIS spectra. 
    Shaded regions during the plateau phase show the light curve interpolation for the \textit{grizJ} bands as described in Sec.~\ref{sec:bolometric}.
    The inset shows early-time photometry, up to 15 days post-explosion, with ATLAS \textit{o} band photometry included (filled circles).
    The dotted line in the inset is the smoothed, interpolated \textit{TESS} light curve.
    The photometry is not corrected for reddening; no offsets are applied except for the $riz$ bands in the inset.
    The offsets used are provided.
    Black and blue ticks on the top of the plot mark the epochs of ground-based and \textit{HST} spectroscopy, respectively.
    The grey dashed line indicates the rate of decline expected from radioactive decay of \cobalt\ in the nebular phase. 
    }
    \label{fig:photometry}
\end{figure*}

SN\,2020fqv was well observed in the \textit{griz} bands with the Panoramic Survey Telescope and Rapid Response System (Pan-STARRS) at Haleakal\={a} Observatory in Hawai`i, as part of YSE. 
In addition, we obtained optical photometry from the Las Cumbres Observatory network \citep{brown2013}, Thacher Observatory (Swift et~al., in prep.), Lulin Observatory, and the Nickel telescope at Lick Observatory.
The optical photometry was processed in the way explained in \cite{kilpatrick2018a} using \texttt{photpipe} \citep{rest2005}.  
We retrieved public photometry of this SN from ATLAS and ZTF \citep{bellm2019, masci2019}.
We also obtained near-IR photometry in the $J$ band using the 0.3-m Gattini-IR telescope at Palomar Observatory \citep{moore2019, de2020}.
\autoref{fig:photometry} shows the light curves of SN\,2020fqv, uncorrected for reddening, along with the interpolated light curves discussed in Sec.~\ref{sec:bolometric}.

\subsection{Swift Photometry}\label{sec:swift-obs}

The Neil Gehrels {\it Swift} Observatory obtained imaging of SN\,2020fqv with the Ultraviolet/Optical Telescope \citep[UVOT;][]{Roming05} from 2020 April 1 to 2020 May 29.  We downloaded these imaging from the calibrated sky frames from the {\it Swift} data archive and performed forced aperture photometry at the site of SN\,2020fqv as determined by aligning the {\it Swift} frames to our Pan-STARRS photometry of the transient described above.  We used {\tt HEASoft} v6.27.2 \citep{heasoft} to perform this analysis with the UVOT aperture photometry method {\tt uvotsource} and an aperture radius of 3\arcsec\ and background radius of 30\arcsec.  All aperture photometry was calibrated using the latest {\it Swift}/UVOT calibration files for {\tt HEASoft}.

\subsection{Ground-based Optical Spectroscopy}\label{sec:optical_spec}
We obtained 20 spectra of SN\,2020fqv ranging from 1 to 373 days post-explosion, well sampling the plateau phase with three epochs in the nebular phase.
\autoref{tab:opt_spec_log} summarizes the observations, providing the epoch of observation, telescope and instrument, and exposure time. 
Spectra presented here were obtained using the SPectrograph for the Rapid Acquisition of Transients (SPRAT; \citealp{piascik2014}) on the Liverpool Telescope (LT); FLOYDS on the 2-meter telescopes of Las Cumbres Observatory (LCO) at Haleakal\={a} Observatory in Hawai`i and Siding Spring Observatory in Australia (\citealp{LCOGT13PASP}); Kast Double Spectrograph \citep{KAST} on the 3-meter Shane telescope at Lick Observatory in California; Gemini Multi-Object Spectrograph (GMOS; \citealp{hook2004}) on the Gemini North Telescope on Maunakea in Hawai`i; and the Low-resolution Imaging Spectrograph (LRIS; \citealp{oke1995}) on the Keck Telescope, also atop Maunakea.   
Spectra were reduced and extracted using standard data reduction pipelines for the respective instruments. 

Six optical spectra were obtained with the GMOS instrument under program GN-2020A-Q-134 (PI: Foley). We used the longslit spectroscopy mode, with the 0.75\arcsec\;slit width and the B600+R400 gratings (wavelength range of 4,000-9,800\;\AA). The spectra were reduced, extracted and calibrated using the IRAF gemini package, with the reduction steps described at the GMOS Data Reduction Cookbook.\footnote{\url{http://ast.noao.edu/sites/default/files/GMOS_Cookbook/}}

Shane/Kast and Keck/LRIS spectra were reduced using our customized data reduction pipeline.\footnote{\url{https://github.com/msiebert1/UCSC_spectral_pipeline}}
It performs the standard field flattening, spectral extraction, wavelength calibration using arc observations, and flux calibration using observations of standard stars taken the same night. A similar procedure was used for the reduction of the LCO/FLOYDS spectra with the pipeline described in \citet{Valenti2014}.

SPRAT spectra where obtained in good seeing (1$"$--1.2$"$ FWHM) conditions. 
They have a wavelength range of 4000-8000 \AA, slit width 1.8 arcsec, R=350 at the centre of the spectrum, with the grating optimized for the blue throughput.
They were reduced and extracted using the SPRAT data reduction pipeline and individual spectra were combined in IRAF with cosmic ray rejection.

\begin{figure*}
    \centering
    \includegraphics[width=\linewidth]{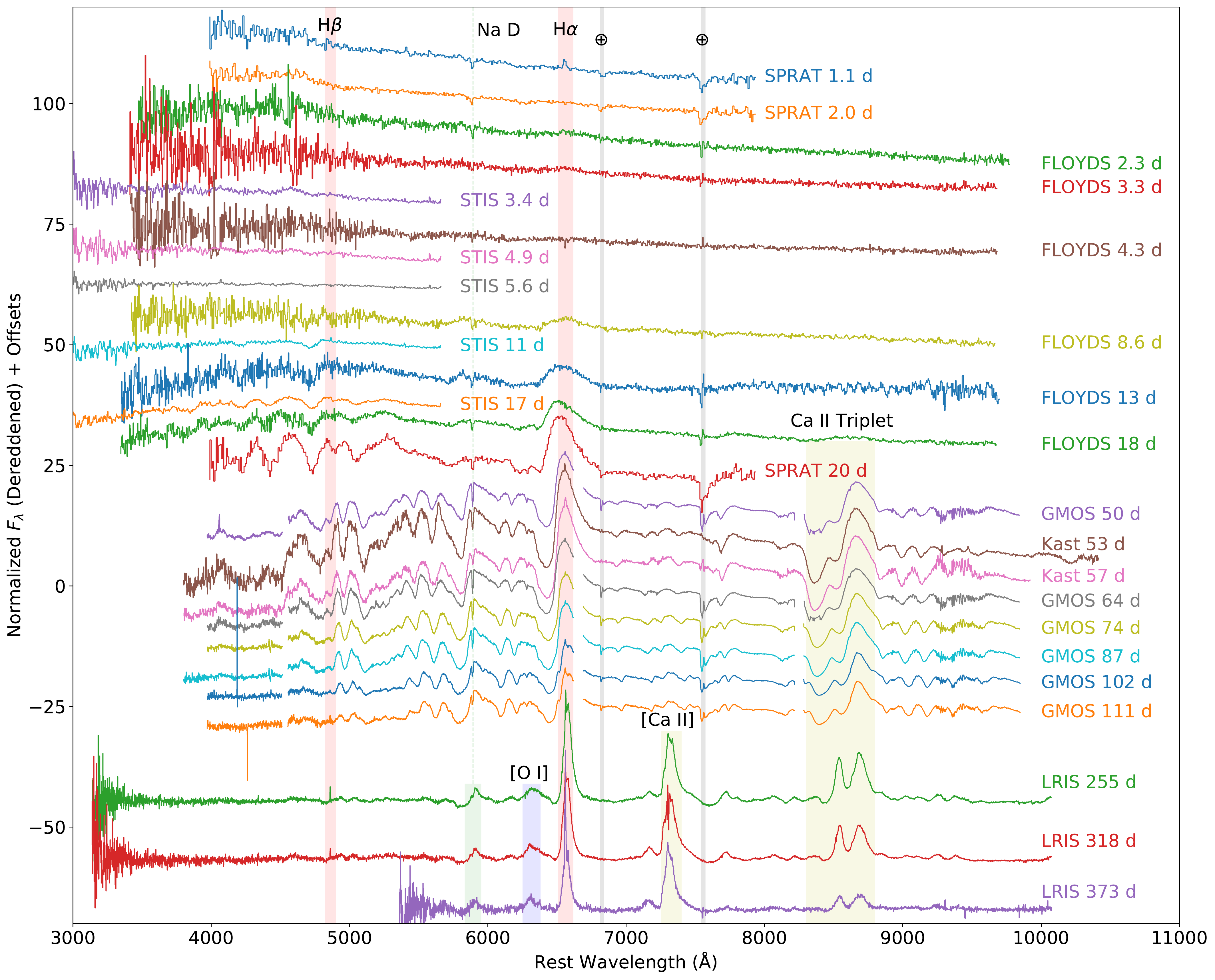}
    \caption{Optical spectra of SN\,2020fqv ranging from 1 to 318 days post-explosion.
    Major spectral lines are marked.
    The H$\alpha$ in the first spectrum at 1.1~d may be from host. 
    The telluric bands are marked with $\oplus$ symbols. 
    The spectra are corrected for dust extinction using the parameters derived in Sec.~\ref{sec:extinction}.
    Note that the GMOS spectra have some gaps in the wavelength coverage due to the chip gaps.
    Also note that the excess flux in the red of the FLOYDS spectrum at 13 d post-explosion with respect to spectra at 8 and 18 d is due to an incomplete galaxy background subtraction.
    The lack of observations between 18 and 50 days post-explosion is due to the shutdown of observatories world wide due to COVID-19. 
    The calibrated spectra uncorrected for reddening are available via WISeREP.}
    \label{fig:optical_spec} 
\end{figure*}

\autoref{fig:optical_spec} shows the sequence of optical spectra of SN\,2020fqv. 
The \ion{Na}{i} D doublet absorption at the host redshift is present in all spectra, and resolved into two lines in GMOS spectra. 
We measured the equivalent width (EW) of the \ion{Na}{i} D 1 and 2 lines to be 1.14 and 2.30, respectively.
The EW of the \ion{Na}{i} D 1 and 2 lines remain constant throughout the plateau phase. 
The Na doublets are saturated at this EW, and their strength is no longer a good probe of the interstellar extinction \citep{poznanski2012}.
We discuss the determination of dust extinction towards this SN in Sec.~\ref{sec:extinction}. 

\begin{table}
\caption{Log of ground-based spectroscopic observations. \label{tab:opt_spec_log}}
\begin{tabular}{lllll}
\hline
Date & MJD   & Epoch & Telescope/Instrument & Exp. Time \\
& & (day) & & (s) \\
\hline
2020-04-01               & 58940.03 & 1.1   & LT/SPRAT            & 900\\
2020-04-01               & 58940.94 & 2.0   & LT/SPRAT            & 600\\

2020-04-02               & 58941.3 & 2.3     & LCO/FLOYDS           & 900                              \\
2020-04-03               & 58942.3 & 3.3     & LCO/FLOYDS           & 900                              \\
2020-04-04               & 58943.3 & 4.3    & LCO/FLOYDS           & 1800                              \\
2020-04-08               & 58947.6 & 8.6     & LCO/FLOYDS           & 1500                             \\
2020-04-13               & 58952 & 13    & LCO/FLOYDS           & 1500                              \\
2020-04-18               & 58957 & 18    & LCO/FLOYDS           & 1500                              \\
2020-04-19               & 58959 & 20    & LT/SPRAT             & 600 \\
2020-05-20               & 58989 & 50    & Gemini/GMOS          & 900                               \\
2020-05-23               & 58992 & 53    & Shane/Kast           & 900                               \\
2020-05-27               & 58996 & 57    & Shane/Kast           & 930                               \\
2020-06-03               & 59003 & 64    & Gemini/GMOS          & 900                               \\
2020-06-13               & 59013 & 74    & Gemini/GMOS          & 900                               \\
2020-06-26               & 59026 & 84    & Gemini/GMOS          & 900                               \\
2020-07-11               & 59041 & 102   & Gemini/GMOS          & 900                               \\
2020-07-20               & 59050 & 111   & Gemini/GMOS          & 1200                              \\
2020-12-11               & 59194 & 255   & Keck/LRIS            & 1800                              \\
2021-02-12               & 59257 & 318   & Keck/LRIS            & 1200 \\
2021-04-08               & 59312 & 373   & Keck/LRIS            & 1500 \\
\hline
\end{tabular}
\end{table}

\subsection{Archival Imaging}

\subsubsection{{\it Hubble Space Telescope}}

We obtained pre-explosion ({\it HST}) imaging of NGC~4568 from the Barbara A. Mikulski Archive for Space Telescopes (MAST)\footnote{\url{https://archive.stsci.edu/hst/}}.  
These data consisted of a single epoch of F606W imaging obtained with the Wide Field Planetary Camera 2 (WFPC2) from 4 Feb 1995 (Program 5446, PI Illingworth), roughly 25 years before the explosion of SN~2020fqv. 
This was the only \textit{HST} image covering the SN site.
We downloaded the calibrated science frames ({\tt c0m}) and aligned, masked, drizzled, and performed {\tt dolphot} photometry \citep{dolphot} using the {\tt hst123} reduction pipeline \citep{Kilpatrick21}.

\begin{figure}
    \centering
    \includegraphics[width=\linewidth]{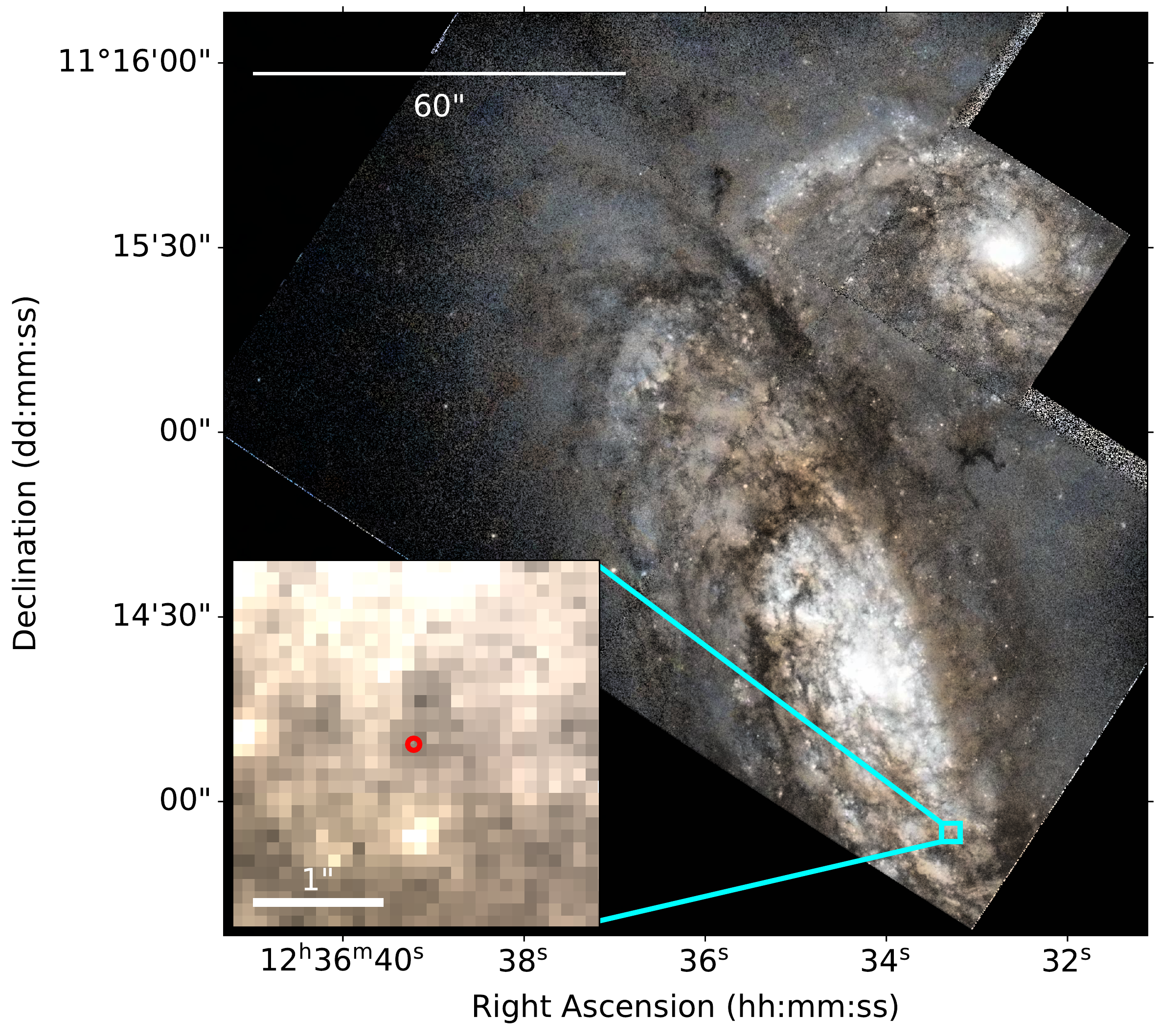}
    \caption{Pre-explosion \textit{HST}/WFPC2 image of NGC\,4567 (northwest) and 4568 (southeast), containing the site of SN\,2020fqv in the F606W filter. 
    The image was taken on 4 Feb 1995 (Program 5446, PI Illingworth).
    We apply square-root stretch to improve the dynamic range of the image. 
    The monochromatic \textit{HST} image is colorized by using the color information from the Digitized Sky Survey (DSS).
    The 3$''\times$5$''$ area around the SN site is shown in an inset with the SN location marked with the red circle. 
    The size of the circle represents the astrometric uncertainty of 0.05$''$ (see text).
    }
    \label{fig:pre_exp_hst}
\end{figure}

We aligned the final drizzled F606W image to a stacked $i$-band frame of SN~2020fqv constructed from our Las Cumbres Observatory follow up imaging.  
Using 5 common astrometric sources, the root-mean square alignment precision was $\sim$0.05\arcsec, or roughly 0.5 WFPC2 pixels. 
\autoref{fig:pre_exp_hst} shows the colorized pre-explosion \textit{HST} image of the SN site with the location of SN\,2020fqv marked. 
There are no sources detected in the WFPC2 image within 6.8 times the astrometric uncertainty. 
We estimated the limiting magnitude in the image by injecting artificial stars with varying magnitudes at the location of SN~2020fqv with varying magnitudes and recovering them with {\tt dolphot}.  In this way, we estimate the 3$\sigma$ limiting magnitude to be $m_{\mathrm{F606W}}=24.80$~mag (AB) as reported in \autoref{tab:archival}.

\subsubsection{{\it Spitzer Space Telescope}}

We downloaded pre-explosion {\it Spitzer Space Telescope} imaging covering the site of SN~2020fqv from the Spitzer Heritage Archive\footnote{\url{https://sha.ipac.caltech.edu/}}.  These consisted of approximately 11,200~s of cumulative exposure in the Infrared Array Camera (IRAC) Channel 1 and 11,100 in Channel 2 and spanning 22 Jun 2004 to 31 Oct 2019, or 15.8 to 0.4~yr before explosion of SN~2020fqv.  Following procedures described in \citet{JacobsonGalan20} and \citet{Kilpatrick21}, we stacked these data using a custom pipeline based on the {\tt photpipe} imaging and photometry pipeline \citep{rest2005}.  The frames for each channel were optimally stacked and regridded to a common pixel scale of 0.3\arcsec~pixel$^{-1}$.  Aligning to our post-explosion imaging, we did not detect any point-like emission at the site of SN~2020fqv, and so injecting artificial stars at this location we recovered limiting magnitudes of 22.74~mag and 23.10~mag on any pre-explosion counterpart to SN~2020fqv in Channel 1 and 2, respectively.

\subsubsection{Pan-STARRS 3$\pi$ Imaging}

We obtained pre-explosion Pan-STARRS imaging in $grizy$ bands from the 3$\pi$ survey \citep{flewelling+16}.  These consisted of stacked frames covering 10.8 to 5.8~yr before the explosion of SN~2020fqv.  Processing these data in {\tt photpipe} as described in \citet{jones2021}, we did not detect any point-like emission at the site of SN~2020fqv.  We derived upper limits on the presence of a counterpart to SN~2020fqv by injecting artificial stars, and we report our 3$\sigma$ limits in \autoref{tab:archival}.

\begin{table}
\caption{Pre-explosion constraints on optical and infrared counterparts to SN~2020fqv.}
    \begin{tabular}{lllll}
\hline
Instrument          & Band & Epoch & Limiting Magnitude \\
                    &      & (day) & (AB mag)  \\ \hline\hline
{\it HST}/WFPC2     & F606W& -9188          & 24.80 \\
PS1/GPC1            & $g$  & -3697 to -2225 & 20.80 \\
PS1/GPC1            & $r$  & -3692 to -2116 & 20.36 \\
PS1/GPC1            & $i$  & -3689 to -2121 & 20.48 \\
PS1/GPC1            & $z$  & -3949 to -2278 & 20.32 \\
PS1/GPC1            & $y$  & -3718 to -2464 & 20.05 \\
{\it Spitzer}/IRAC  & Ch1  & -5780 to -158  & 22.74 \\
{\it Spitzer}/IRAC  & Ch2  & -5780 to -158  & 23.10 \\ \hline
    \end{tabular}
    \label{tab:archival}
\end{table}

\section{Analysis}\label{sec:analysis}
\subsection{Dust Extinction}\label{sec:extinction}
There is substantial dust extinction in the line of sight towards SN\,2020fqv. 
The lack of the Na I D absorption at $z = 0$ and the low value of Galactic extinction of $E(B-V)_{\mathrm{MW}} = 0.029$ mag \citep{schlafly2011} indicate that the extinction is primarily from the host galaxy. 
We estimate the dust extinction by assuming that at 4 d post-explosion, the spectrum of the SN is well described by a black body. 
We then fit the \textit{HST} and ground-based spectra from 4 d with a black body model, leaving the temperature, black body radius, and the dust extinction parameters $E(B-V)$ and $R_V$ as free parameters. 
The extinction law used is \cite{fitzpatrick1999}.
The fit is performed using the Markov Chain Monte Carlo (MCMC) package \texttt{emcee} \citep{foreman2013}. 
The prior distribution for all parameters is uniform for values that are physical (e.g., positive temperature and radius).
The best-fit extinction parameters are $E(B-V) = 0.52 \pm 0.01$ mag and $R_V = 3.19 \pm 0.04$. 
\autoref{fig:dust_fit} shows the result of this fit. 
We use these extinction corrections for all subsequent analyses.

\begin{figure}
    \centering
    \includegraphics[width = \linewidth]{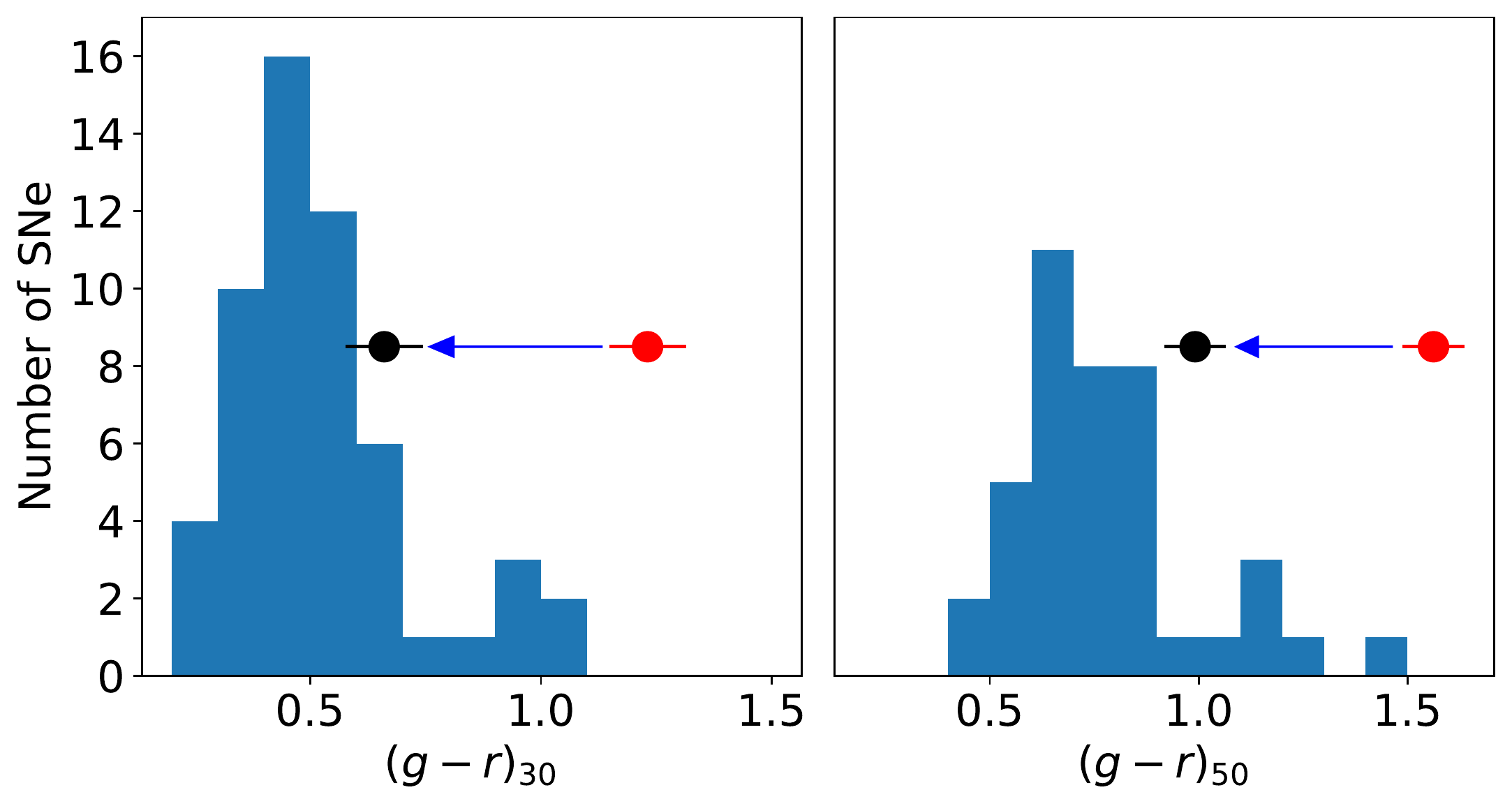}
    \caption{
    Histograms showing the $g-r$ color of SNe II from \protect\cite{dejaeger2018} at 30 (\textit{left}) and 50 (\textit{right}) days post-explosion. The colors of SN\,2020fqv before and after the extinction correction are plotted in red and black, respectively, in each plot.
    This demonstrates that after the extinction correction derived from early-time spectroscopy, the color of SN\,2020fqv is similar to that of other SNe II.
    }
    \label{fig:color_correction}
\end{figure}

To check that the reddening correction is reasonable, we compare the $g-r$ color of SN\,2020fqv at 30 and 50 d post-explosion, $(g-r)_{30, 50}$, with a sample of SNe II from \cite{dejaeger2018}. 
Fig.~\ref{fig:color_correction} shows histograms of the $g-r$ colors at 30 and 50 days post-explosion from \cite{dejaeger2018}, and the colors of SN\,2020fqv before and after the correction.
Before we correct for the dust reddening from the host galaxy, the observed colors of SN\,2020fqv are $(g-r)_{30, \rm obs} = 1.23 \pm 0.08$ mag and $(g-r)_{50, \rm obs} = 1.56 \pm 0.07$ mag, well outside of the color distribution of SNe II in \cite{dejaeger2018} ( Figs.~A3 and A4 {in their paper}).
With our inferred dust reddening parameter, the $g-r$ color correction is 0.57 mag, which brings the colors down to $(g-r)_{30} = 0.66 \pm 0.08$ mag and $(g-r)_{50} = 0.99 \pm 0.07$ mag.
These colors are consistent with other SNe II in the sample.

\subsection{Time of explosion}\label{sec:exp_epoch}
With early \textit{TESS}, ATLAS, and ZTF $g$ and $r$ photometry, we can determine the explosion time by fitting analytical CCSNe rise models. We fit the RSG model, without shock breakout, from \citet{Nakar2010} following \citet{Garnavich2016} and \citet{vallely2021}. We set the progenitor mass to be $15~ M_\odot$, allowing progenitor radius, explosion energy and explosion time to be free parameters. 
We will show how we derive this mass estimate from light curve fitting (Sec.~\ref{sec:scaling}, \ref{sec:LC_model}) and nebular spectroscopy (Sec.~\ref{sec:nebular_spec}.
Since the Nakar models assume black body emission, we apply the extinction derived in Sec.~\ref{sec:extinction} to the black body spectrum using the extinction law of \citet{fitzpatrick1999}. Magnitudes are then calculated by applying the \textit{TESS}, ATLAS $o$, and ZTF bandpasses to the spectrum and integrating over it at each time step.

We simultaneously fit all free parameters to \textit{TESS}, ATLAS, and ZTF photometry before $\rm MJD = 58957$ using \texttt{emcee} \citep{Foreman-Mackey2013}.
We use flat priors for all variables and maximise $-\chi^2$ in flux space. The best fit parameters, shown in \autoref{fig:nakar_fit}, are $R=120^{+40}_{-30}\rm ~R_\odot$, $E=0.77^{+0.17}_{-0.15}\times 10 ^{51}$~erg, and explosion time of $t_0 = 58938.93^{+0.15}_{-0.16}$~MJD,
where $R$ and $E$ are the progenitor radius and the explosion energy, respectively. 
We note that these values are different from those presented in \citet{vallely2021} for SN\,2020fqv, however, there are a few key differences. For example, we use a photometric zeropoint calibrated to concurrent spectra and photometry, this will lead to different peak fluxes. Furthermore, while \citet{vallely2021} fit to an analytical model, we fit the semi-analytic \citet{Nakar2010} model simultaneously to multiple bands. We also note that the radius is small for a RSG, but this is because the model does not take CSM interaction into account. 

Although we constrain progenitor properties with this model, aside from explosion time, they are unreliable. As can be seen in \autoref{fig:nakar_fit}, the best fit model can not match the rapid rise of the SN~2020fqv light curves in all bands, likely due to ejecta interacting with a dense circumstellar medium. Despite the shortcomings, this model provides us with a self-consistent and physically motivated way to extrapolate all light curves to early times. While, the fit fails to produce realistic parameters for the progenitor radius, it does agree well with early observations in all bands until $\sim58940$~MJD. Therefore, we only use the explosion time from this fit, and model progenitor properties in Sec.~\ref{sec:progenitor}.

\subsection{Quasi-bolometric luminosity}\label{sec:bolometric}
To estimate the total radiative energy, we compute the quasi-bolometric luminosity of SN\,2020fqv using the following steps. 
First, the photometric data in different bands are generally taken at different epochs, so we interpolate them onto a common time grid. 
We only perform this interpolation for the data during the plateau phase.
We use the Gaussian process regression package \texttt{george} to perform the interpolation \citep{ambikasaran2015}.
We use the Mat\'{e}rn 3/2 kernel, which is less likely to smooth out features of the light curve compared to the more commonly used exponential squared kernel as suggested in \cite{boone2019}. 
The length-scale parameter used for the Gaussian process was 200 d.
We find no significant difference between using the two kernels. 
We also find no difference in results between interpolating each band at a time and interpolating all bands simultaneously, assuming that the evolution in one band resembled that in the adjacent bands. 
We do not extrapolate the data more than two days from the actual observation.
The interpolation is performed in the flux space.
Along with the photometry, \autoref{fig:photometry} shows the interpolated light curves in each band.

\begin{figure}
    \centering
    \includegraphics[width = \linewidth]{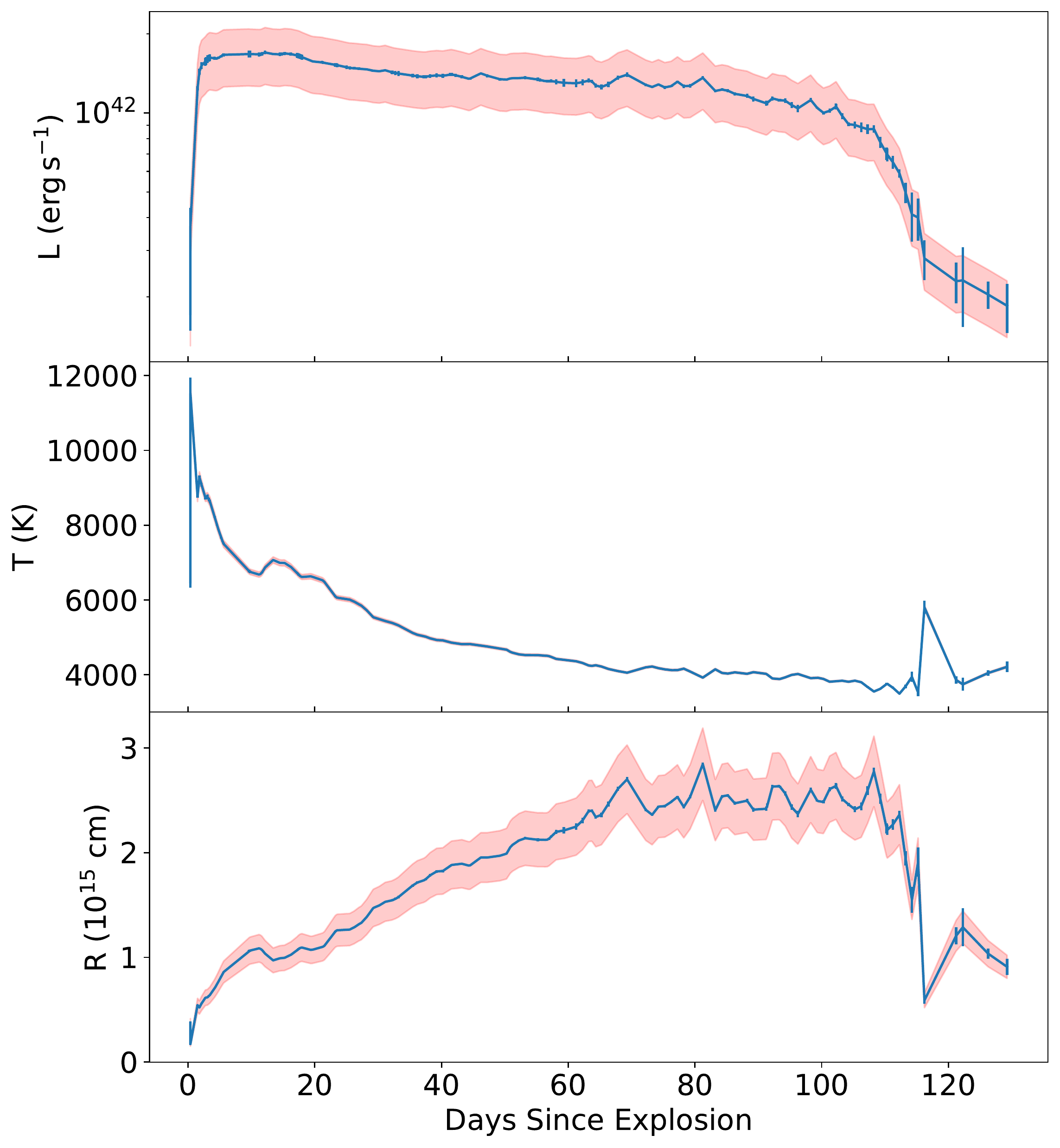}
    \caption{Results of fitting a black body model to the multi-band light curves of SN\,2020fqv. From top to bottom: quasi-bolometric luminosity, black body temperature, and black body radius. For each of the black body parameters, error bars represent 1-$\sigma$ statistical uncertainty due to photometric uncertainties; shaded bands represent the systematic uncertainty from the extinction correction and the distance to the SN. 
    For the luminosity and black body radius, the systematic uncertainty is dominated by the distance uncertainty.
    For the temperature, only the uncertainty in the extinction correction contributes to the systematic uncertainty. 
    }
    \label{fig:bb_fit}
\end{figure}

For each epoch, we fit the multiband photometry with a black body model reddened by the dust extinction model discussed in Sec. \ref{sec:extinction}.
For epochs with more than two bands available, we do not fit the $r$~band to avoid the significant H$\alpha$ line. 
\autoref{fig:bb_fit} shows the resulting luminosity, temperature, and black body radius.
The error bars in the plot represent statistical errors due to photometric uncertainties. 
The shaded bands represent systematic errors due to the uncertainty in the extinction correction and the distance to the SN. 
Note that for the luminosity and radius, the distance uncertainty dominates the overall uncertainty; for the temperature, only the extinction uncertainty contributes to the systematics. 


\subsection{Progenitor mass and explosion properties} \label{sec:progenitor}
We estimate the explosion properties of SN\,2020fqv, which are the explosion energy, progenitor mass, and \nickel\ mass, using four independent methods. 

\subsubsection{Scaling relations derived from models}\label{sec:scaling}
First, we obtain estimates of the explosion properties by using scaling relations from \cite{sukhbold2016}, which are derived using results from the \texttt{KEPLER} SN explosion models.
Equations 15 and 19 from \cite{sukhbold2016} relate the explosion energy, hydrogen envelope mass, and progenitor radius to the observed parameters: luminosity at 50 days ($L_{50}$), plateau length ($t_p$), and the amount of $^{56}$Ni synthesized ($M_{\rm Ni}$).
The luminosity at 50 days can be obtained directly from our quasi-bolometric light curve described in the last section: $L_{50} = (1.3 \pm 0.3) \times 10^{42} \, \rm erg\, s^{-1}$.
To measure the plateau length, we follow \cite{valenti2016} and fit their Equation 1 to the light curve around the transition between the plateau and the nebular phase; we find the plateau length of $114 \pm 1$ d.
We measure the amount of \nickel\ synthesized in SN\,2020fqv by measuring its luminosity in the radioactive decay tail. 
The only epoch where multi-band photometry exists for this purpose is at 270 d post-explosion.
At that epoch, SN\,2020fqv's luminosity from photometry is $(5.6\pm 2.1) \times 10^{40} \, \rm erg\, s^{-1}$ (the uncertainty includes both uncertainties from photometry and the distance to the SN). 
Because the decline rate in this phase follows what is expected from radioactive decay of \cobalt, it is reasonable to assume that gamma ray from radioactivity is efficiently absorbed and reemitted in the optical. 
With this assumption, the luminosity of the SN reflects the heating rate from the radioactive decay of \cobalt\ at the time, and we can compute the amount of \cobalt\ present at that epoch: $M_{\rm Co}({\rm 270 \, d}) = L({\rm 270 \, d}) / \epsilon_{\rm Co}$ where $\epsilon_{\rm Co} = 6.8 \times 10^9 \rm \, erg\, s^{-1} \, g^{-1}$ is the heating rate from the \cobalt\ decay. 
Then the \nickel\ mass synthesized (at $t = 0$) is:
\begin{equation}
    M_{\rm Ni} = M_{\rm Co}(t) \frac{\lambda_{\rm Co} - \lambda_{\rm Ni}}{\lambda_{\rm Ni}}\left[ \exp{( - \lambda_{\rm Ni} t)} - \exp{(-\lambda_{\rm Co} t)} \right]^{-1}
\end{equation}
where $\lambda_{\rm Ni, Co}$ are the inverse of the radioactive decay timescale of \nickel\ and \cobalt, respectively.
With this calculation, we infer $0.043\pm 0.017 \, M_{\odot}$ of $^{56}$Ni synthesized in SN\,2020fqv. 
To double check, we also compare the luminosity of SN\,2020fqv to that of SN\,1987A, which is known to produce 0.075 \msun of \nickel\ following e.g., \cite{spiro2014}.
At 270 d post-explosion, $L_{\rm 87A} = 9.0 \times 10^{40} \, \rm erg\, s^{-1}$ \citep{arnett1989}, and we arrive at $M_{\rm Ni} \approx 0.047$ \msun\, for SN\,2020fqv, consistent with the first calculation. 
We adopt 0.043 \msun as the mass of \nickel\ synthesized in SN\,2020fqv. 

Lastly, to break the degeneracy between the hydrogen envelope mass and the progenitor radius, we consider the values listed in Table 2 in \cite{sukhbold2016} where $M_i$ is the initial mass, $M_f - M_\alpha$ is the envelope mass, and $R_f$ is the progenitor radius. 
The relationship between $M_f - M_\alpha$ and $R_f$ is monotonic for $M_i < 20 \, M_\odot$, so we only use the values from this range.
Other methods discussed later show that this assumption is reasonable.

With these quantities determined, we solve the aforementioned Equations 15 and 19 from \cite{sukhbold2016} using the \texttt{fsolve} function in the \texttt{scipy.optimize} package.
We arrive at the explosion energy and progenitor mass of $(4.1 \pm 0.1) \times 10^{50} \, \rm erg$ and $15 \pm 3 \, M_\odot$, respectively. 
For these values, the progenitor radius and the hydrogen envelope mass are $800 \pm 100 \, R_\odot$ and $8.5 \pm 0.7 \, M_\odot$, respectively.


\subsubsection{Comparison with numerical models}\label{sec:LC_model}
\begin{figure}
    \centering
    \includegraphics[width=0.9\linewidth]{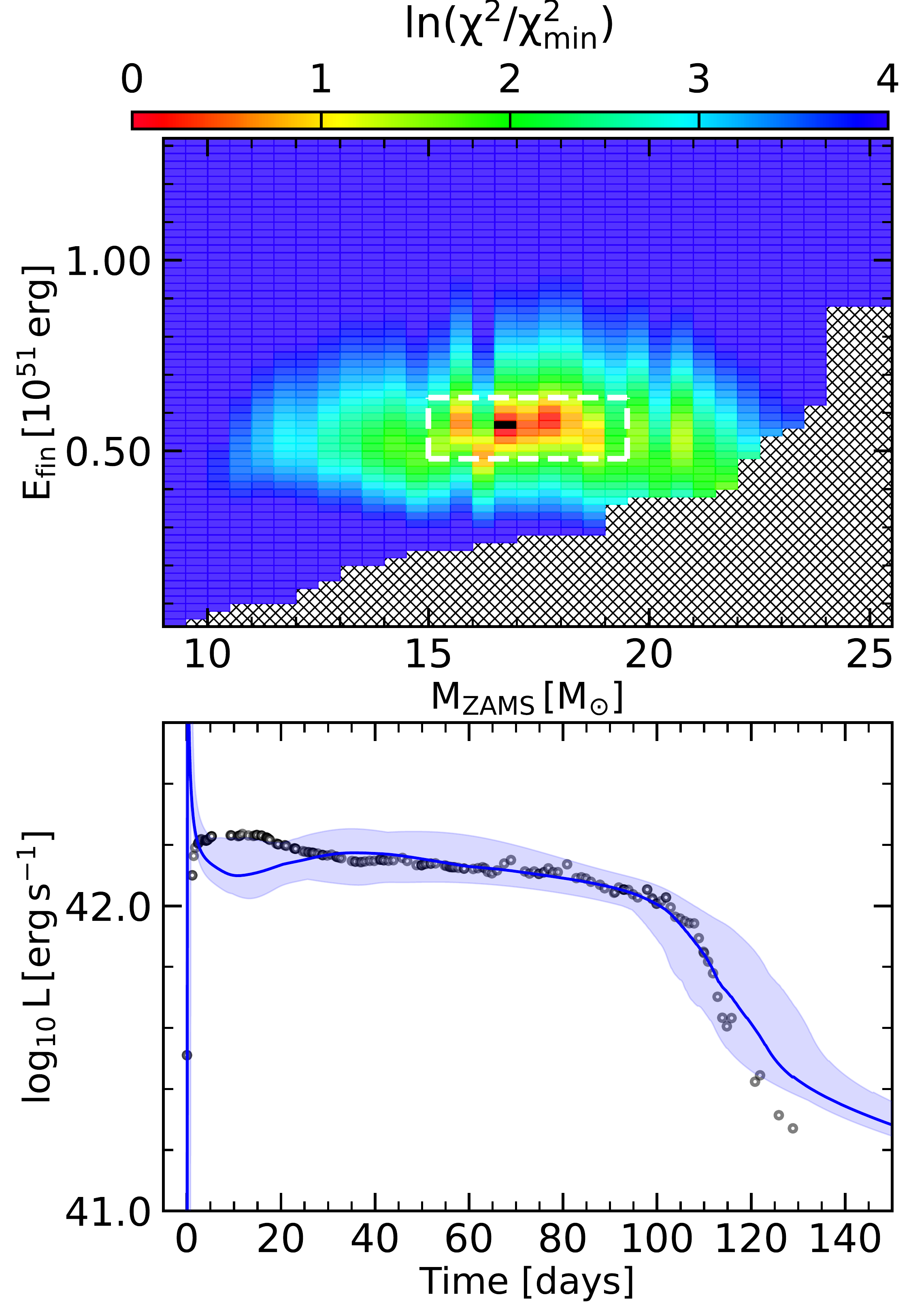}
    \caption{Top: A heat map of the goodness-of-fit figure $\ln(\chi^2/\chi^2_{\rm min})$ of different SNEC models compared to the data as a function of the zero-age main sequence mass (ZAMS) of the progenitor star and the kinetic energy of the explosion.
    The hatched region indicates the parameters for which we could not obtain theoretical light curves because of fallback material.
    The result shows that the best-fit explosion parameters of SN\,2020fqv are $M_{\rm ZAMS} = 16.5 \pm 1.5 \, M_\odot$ and $E = 5.5 \pm 0.7 \times 10^{50} \, \rm erg$.
    This uncertainty box is indicated as a white box in the top panel.
    These parameters are in rough agreement with the values we derived from simple scaling relation.
    Bottom: The best-fit model of the bolometric luminosity of SN\,2020fqv. Note that the CSM-less model underpredicts the luminosity less than one month post-explosion.
    The shaded region represents the models from the uncertainty box in the top panel.}
    \label{fig:no_csm_fit}
\end{figure}

In order to independently confirm the explosion parameters, we compared our observations with a grid of models generated using the SuperNova Explosion Code (SNEC; \citealp{morozova2015}) for SN\,2004et, which produced a similar amount of $^{56}$Ni \citep{morozova2018}.
These models do not account for CSM interactions.
\autoref{fig:no_csm_fit} (top) shows the heat map of the goodness-of-fit figure for the models with different kinetic energies and progenitor masses, showing that the best-fit values are $E = (5.5 \pm 0.7) \times 10^{50} \, \rm erg$ and $M_{\rm ZAMS} = 16.5 \pm 1.5 \, M_\odot$, respectively. 
These numbers are roughly consistent with what we derived from scaling relations in Section \ref{sec:scaling}. 
\autoref{fig:no_csm_fit} (bottom) shows the best-fit light curve on top of the quasi-bolometric luminosity from \autoref{fig:bb_fit}. 
We note that at early time prior to 35 days post-explosion, the model substantially underpredicts the luminosity. 
In the next section, we perform a similar fit accounting for the extra flux from CSM interaction to fully explain the light curve. 
One consequence from the CSM interaction is that it prolongs the plateau. 
As such, the progenitor mass required to explain the plateau mass is smaller compared with the CSM-free models used in this section by about 1.5 \msun.
Thus, the range of $M_{\rm ZAMS}$ from this method is $15 \pm 1.5$ \msun.

\subsubsection{Late-time nebular spectroscopy}\label{sec:nebular_spec}
As an independent measure of the progenitor mass of SN\,2020fqv, we compare the nebular phase spectra obtained at 255, 318, and 373 days post-explosion with models from \cite{jerkstrand2014} (12--25 \msun; high mass) and \cite{jerkstrand2018} (9 \msun, low mass). 
Similar approach has been used in the literature to constrain the progenitor mass of SNe~II \citep[e.g.,][]{terreran2016, silverman2017, vandyk2019,bostroem2019, szalai2019, hiramatsu2021}.
The high-mass models in \cite{jerkstrand2014} were computed for SN\,2012aw at a distance of 5.5 Mpc and with a \nickel\ mass of 0.062 \msun. 
The low-mass models in \cite{jerkstrand2018} were computed for a SN from a 9 \msun\, progenitor star at a distance of 10 Mpc with a \nickel\ mass of 0.0062 \msun.
We scale the flux proportionally with the \nickel\ mass and inverse-proportionally with distance squared.
We note that for the 9 \msun\, models, the scaling factor due to the \nickel\ mass is $\sim$70.
The closest epochs to the data for which high-mass models are available are 250, 306, and 369 days (for the second epoch, the closest 19 \msun\, model available is at 332 d; for the third epoch, the closest 12 \msun\, model available is at 400 d). 
The closest epochs to the data for which low-mass models are available are 200, 300, and 400 days. 
To account for these differences in epoch, we scale the flux with a factor of $\exp(x/111.3)$ where x is the difference between the epoch of the model and the data in days and 111.3 is the lifetime of the \cobalt\ decay. 
For instance, for the data at 255 d, the models for 250 d are scaled by a factor of $\exp(-5/111.3)$. 

The data must be flux calibrated to be compared to the model. 
To do so, we compute synthetic photometry of the spectra in the $r$ band by computing 
\begin{equation}
    \int F_\lambda (\lambda) T(\lambda) \, \mathrm{d}\lambda / \int T(\lambda) \, \mathrm{d}\lambda 
\end{equation}
where $F_\lambda$ is the flux spectrum and $T(\lambda)$ is the filter transmission profile. 
We then scale to spectrum to match the synthetic photometry with the observed photometry interpolated to the spectrum's epoch. 
We account for dust extinction in both the photometry and the spectra.

\begin{figure*}
    \centering
    \includegraphics[width = 0.8\textwidth]{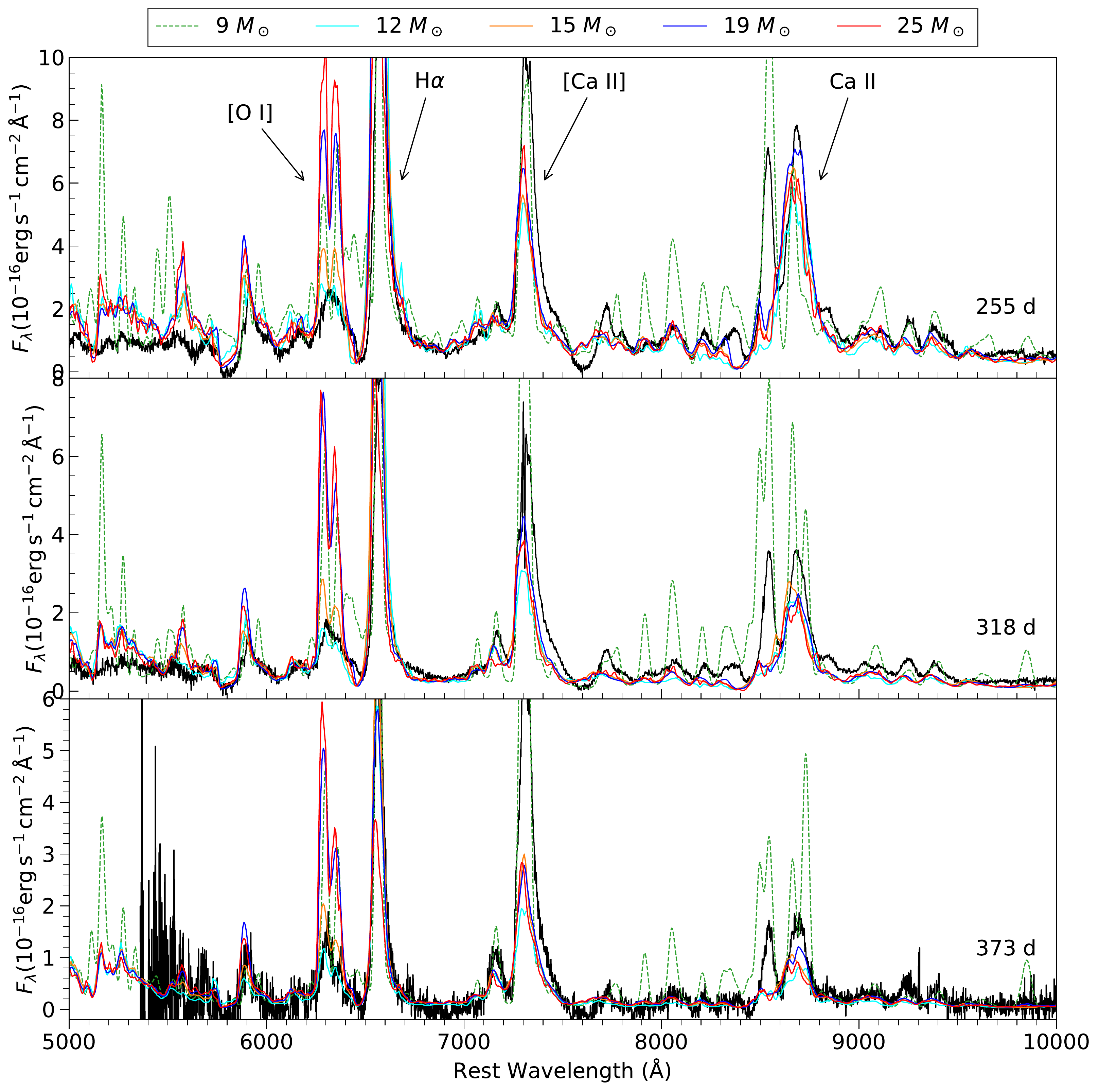}
    \caption{Nebular spectra of SN\,2020fqv at 255 (top), 318 (middle), and 373 (bottom) days post-explosion. 
    Flux calibration is obtained by comparing the synthetic photometry from these spectra with the interpolated observed $r$ band photometry.
    Nebular spectra models from \protect\cite{jerkstrand2014} for $\protect M_{\mathrm{ZAMS}} = 12$, 15, 19, and 25 \msun\, and the 9 \msun\, models from \protect\cite{jerkstrand2018} are overplotted for comparison. 
    Significant emission lines ([O I], H$\alpha$, [Ca II], and Ca II triplet) are annotated. 
    See Section \ref{sec:nebular_spec} for how these models have been scaled to match SN\,2020fqv.
    The 12 and 15 \msun\,models provide the best fit to our data.
    }
    \label{fig:nebular_spec}
\end{figure*}

\autoref{fig:nebular_spec} shows the spectra with models overplotted. 
The models at four different masses generally predict similar line luminosity for the calcium lines: both the IR triplet and the [Ca \textsc{II}] 7292, 7324 \AA. 
We note that the discrepancy between the model and the IR triplet profile of SN\,2012aw has been noted in the original study \citep{jerkstrand2014}. 
The main discriminating feature between the four models is the luminosity of the [O \textsc{I}] 6300, 6364 \AA\ lines, since the oxygen mass has been shown to be proportional to the ZAMS mass \citep[e.g.,][]{dessart2011, jerkstrand2014}. 
From the three epochs of data, models with $M_{\rm ZAMS} = 12$ and 15 \msun\, are favored over the 19 and 25 \msun\, models. 
The low-mass 9 \msun\, models predict stronger line emissions than what is observed. 
This is likely due to the fact that the models have been scaled up by a large factor to make up for the small \nickel\ mass assumed in the model. 
The inferred progenitor mass is consistent with the progenitor mass inferred from light curve fitting. 
We also note that the models predict the continuum flux level of SN\,2020fqv well, especially in the blue, showing that our reddening correction is reasonable. 
From these independent measurements, we conclude that the progenitor star of SN\,2020fqv is an average mass RSG with $M_{\rm ZAMS}$ around 12--15 \msun.

\subsubsection{Progenitor mass limit from pre-explosion imaging non-detection} \label{sec:non_detection}

The flux upper limits of a pre-explosion counterpart to SN~2020fqv from \autoref{tab:archival} allow us to constrain the properties of the progenitor star of this SN.
Assuming consistent Milky Way extinction and interstellar host extinction as in Sec.~\ref{sec:extinction}, we model hypothetical counterparts to SN~2020fqv as blackbodies with a fixed intrinsic luminosity and temperature.  
Following methods described in \citet{Kilpatrick21}, we then forward model this spectral energy distribution assuming a host extinction, redshift, distance, Milky Way extinction, and filter transmission functions consistent with our previous analysis.  
If any of the derived magnitudes in {\it HST}/WFPC2 F606W, {\it Spitzer}/IRAC Channels 1 and 2, or PS1/GPC1 $grizy$ are brighter than those given in \autoref{tab:archival}, we consider that model ruled out.

\begin{figure}
    \centering
    \includegraphics[width=\linewidth]{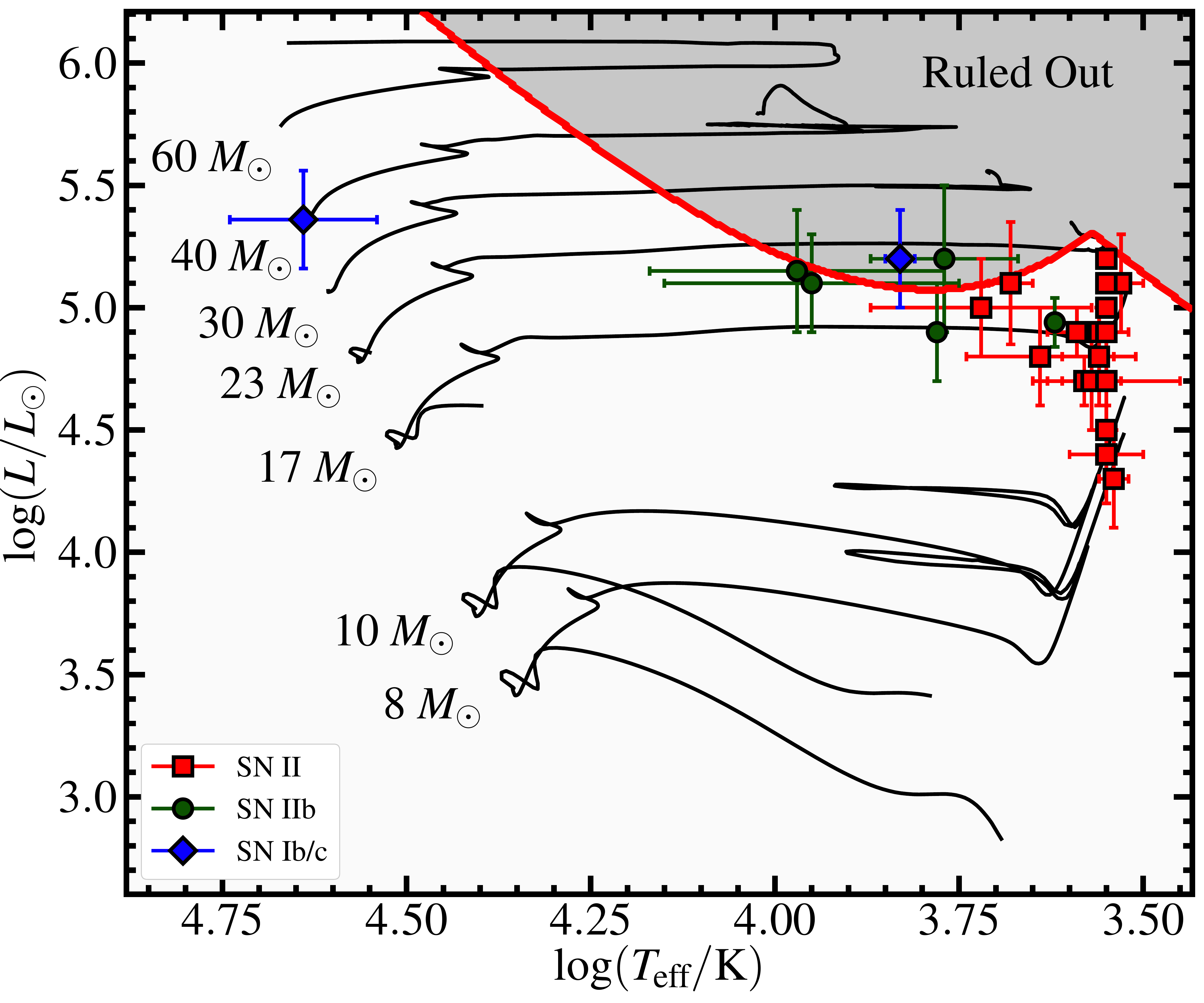}
    \caption{Hertzsprung-Russell diagram showing the parameter space for which we rule out a pre-explosion counterpart to SN~2020fqv using the limiting magnitudes in \autoref{tab:archival}.  The grey region is ruled out while stars occupying the remainder of the diagram are allowed.  For comparison, we show SN progenitor stars including red supergiant progenitor stars to Type II SNe \citep{Smartt15}, red and yellow supergiant progenitors to Type IIb SNe \citep{aldering+94,crockett+08,maund+11,vandyk+14,kilpatrick17:16gkg}, and the progenitor systems of the Type Ib SNe iPTF13bvn \citep{Cao13} and SN~2019yvr \citep{Kilpatrick21}.  We also overplot single-star evolutionary tracks from the MESA Isochrones \& Stellar Tracks code \citep{Choi16}.}
    \label{fig:preexplosion}
\end{figure}

The results of this analysis are shown in \autoref{fig:preexplosion}.  
We can rule out most evolved and terminal massive stars with $\log(L/L_{\odot})>5.1$, including the yellow supergiant progenitors to SN~2008ax and SN~2019yvr \citep{crockett+08,Kilpatrick21}.  
However, most known RSG progenitor stars are allowed by our limits, including all verified RSG detections in \citet{smartt2009} and \citet{Smartt15}.  
Quantitatively, our limits correspond to $\log(L/L_{\odot})\approx5.1$ along the RSG branch where $T_{\mathrm{eff}}\approx3450$~K.  
Thus following single-star evolutionary tracks from the MESA Isochrones \& Stellar Tracks code \citep[MIST;][]{Choi16}, we find that our limits rule out red supergiant progenitors with $M_{\mathrm{ZAMS}}>17~M_{\odot}$.

\subsection{Circumstellar medium interaction} \label{sec:csm_model}

\subsubsection{Early excess flux from CSM interaction}

In this section we attempt to explain the early excess flux
observed in SN\,2020fqv, assuming that the CSM 
around the progenitor originated from an outburst caused
by wave heating during late-stage nuclear burning \citep{quataert2012, fuller2017, wu2021}. 
We note that the mechanism of CSM formation has not
been established yet, and different scenarios are possible, 
but this approach proved fruitful in our earlier work on 
SN 2017eaw \citep{morozova2020}. In this approach, we first pre-heat a
RSG model by launching a weak shock wave through its 
envelope, which leads to the ejection of outermost material.
As this material expands, we collect the snapshots of its profile, which extends from the stellar surface out to some outer radius. 
We later model the interaction between the SN shock with this CSM.

Both initial pre-heating of the RSG star and the final SN explosion are 
modeled with the publicly available code \texttt{SNEC}
\citep{morozova2015}. We work with a solar metallicity, $15\, M_{\odot}$
(at zero-age main sequence, ZAMS) stellar evolution model
from the \texttt{KEPLER} set by \citet{sukhbold2016}, evolved to the
pre-collapse RSG stage. To pre-heat the star we inject 
energy $E_{\rm inj}$ at the base of its hydrogen
envelope with the values between $1.0\times 10^{46}$ and
$10.5\times 10^{46}\,{\rm erg}$, in steps of $0.5\times 10^{46}\,{\rm erg}$.
These values of $E_{\rm inj}$ are chosen based on the models of vigorous
late-stage nuclear burning episodes (core Ne or O burning) 
described {for a similar solar
metallicity $15\,M_{\odot}$ \texttt{MESA} model} in \citet{fuller2017}.
The snapshots of pre-heated density profiles 
are collected every $\sim 20$ days while we track the dynamics
of the resulting outburst until $\approx900$ days after the energy
injection. Finally, the obtained grid of models in $E_{\rm inj}$ and
$t_{\rm inj}$ (time since the energy injection) is exploded in a 
regular core-collapse SN setup 
with final energy $E_{\rm fin}=0.5\times 10^{51}\,{\rm erg}$
and radioactive $^{\rm 56}$Ni  mass $M_{\rm Ni}=0.043\,M_{\odot}$ mixed
uniformly up to $7\,M_{\odot}$ in mass coordinate.

In addition to bolometric luminosity, \texttt{SNEC} outputs color
light curves computed from a black body spectrum taken
at the photospheric temperature. At early times (first $\sim 30$
days of the light curve), when the photospheric temperature
exceed $\sim 8000\,{\rm K}$, the SN spectra are 
sufficiently close to the black
body spectra. However, as the temperature of the ejecta decreases
to $\sim 6500\,{\rm K}$ (plateau part), the blue part of the spectrum
becomes affected by the blanketing from large number of 
iron group lines (see \citealt{kasen2009}). To correct our light
curves for this effect we post-process the output of SNEC with the
spectral code \texttt{SYNOW} \citep{thomas2011}\footnote{https://c3.lbl.gov/es/}.
Namely, we construct \texttt{SYNOW} spectra using ejecta velocities and
temperatures returned by SNEC, and obtain the color light
curves in different observed bands from these spectra 
using the filter profiles. The best-fit model is determined from comparing
the obtained color light curves to the observed magnitudes of
SN\,2020fqv and looking for the minimum of $\chi^2$.

\begin{figure}
  \centering
  \includegraphics[width=0.445\textwidth]{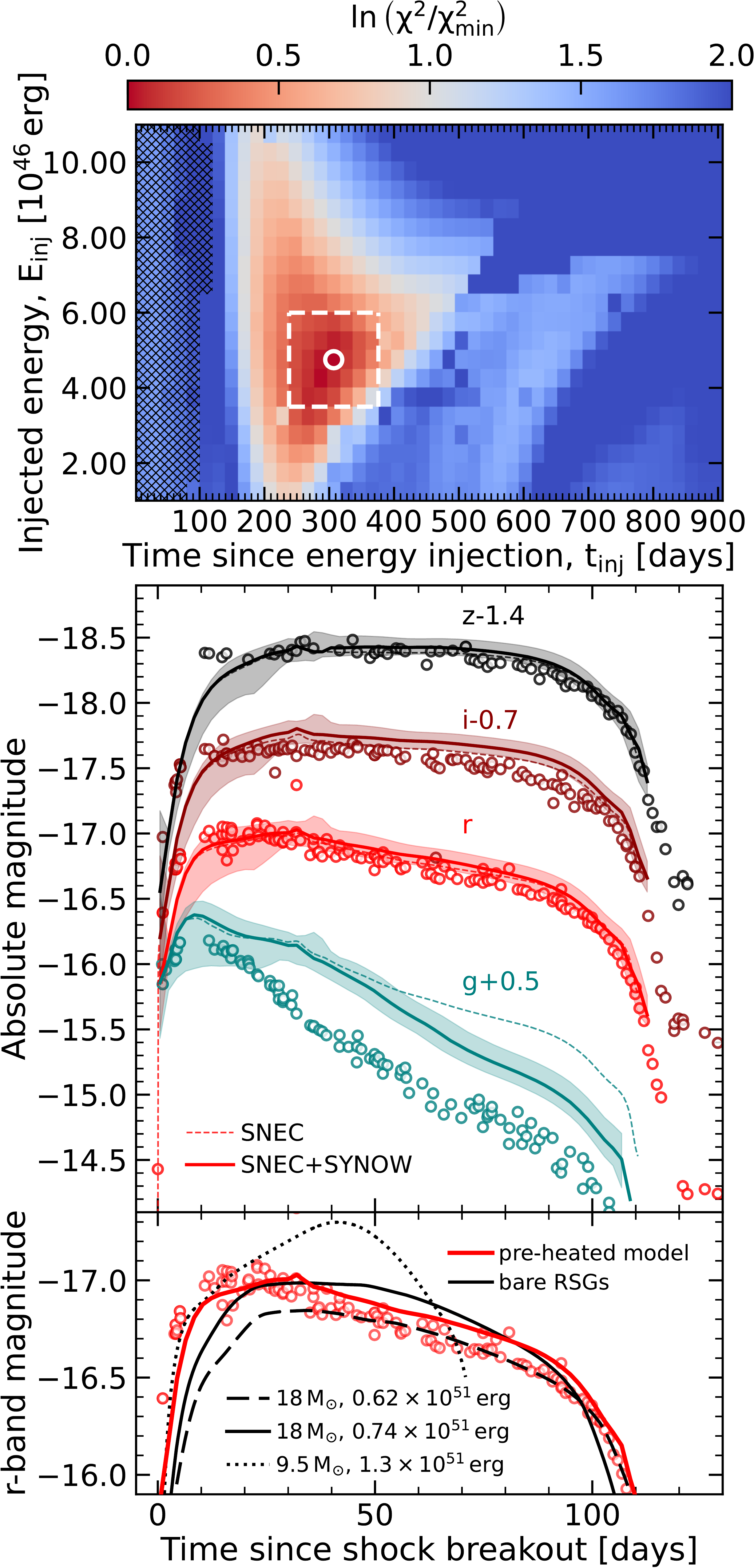}
  \caption{{\bf Top panel:} Color coded distribution of 
  $\ln\left(\chi^2/\chi_{\rm min}^2\right)$
  across the $E_{\rm inj}$-$t_{\rm inj}$ parameter space. The black cross-hatched region covers the times during which the weak shock wave
  launched by the energy injection has not yet reached the stellar surface.
  All final SN explosions are performed with the same set of parameters,
  $E_{\rm fin}=0.5\times10^{51}\,{\rm erg}$ and $M_{\rm Ni}=0.043\,M_{\odot}$.
  The white circle shows the minimum $\chi^2$, while the dashed white
  rectangle outlines the models that were used to construct the error bars in
  the bottom panel. {\bf Middle panel:} The color light curves of the best fitting
  model, compared to the data of SN\,2020fqv. Dashed lines show the magnitudes
  returned by SNEC, while the solid curves are obtained by post-processing
  SNEC output with SYNOW. The error bars correspond to the area between
  all light curves within the white rectangle in top panel. {\bf Bottom panel:} {The $r$-band light curve of our best fitting model (solid
  red line) together with the data of SN 2020fqv. For comparison, the
  black lines show our best attempts to fit the $r$-band data with the bare RSG
  models that were not pre-heated by a weak shock wave before the full
  SN explosion. The dashed, solid, and dotted black curves aim to fit
  the data in the time intervals $35-110$ days, $0-110$ days, and $0-20$ days,
  respectively. 
  } } 
  \label{fig:color_fits}
\end{figure}

\begin{figure}
    \centering
    \includegraphics[width=\linewidth]{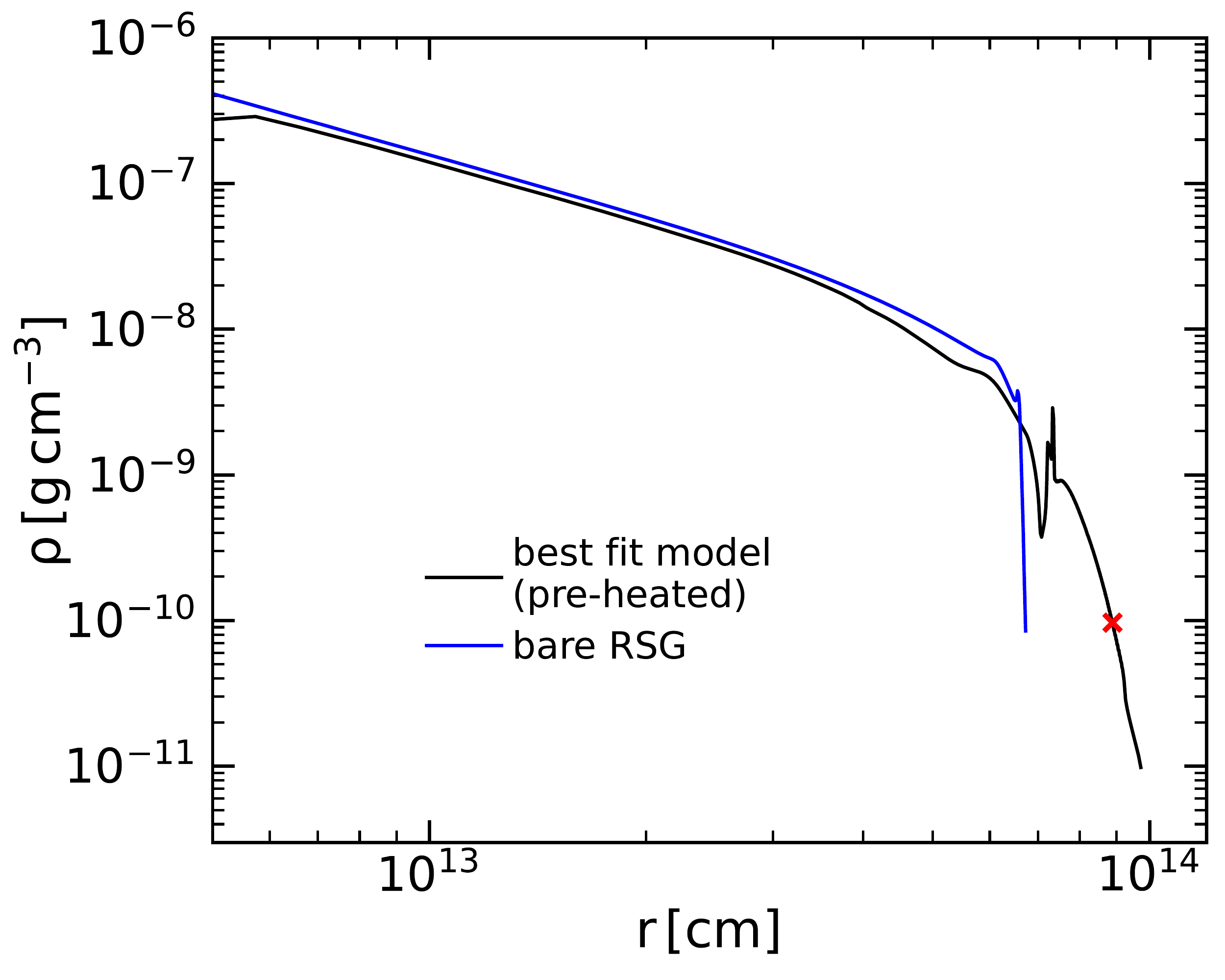}
    \caption{The density profile of the progenitor star of SN\,2020fqv and its CSM at the time of explosion. 
    This density profile is used in \autoref{fig:color_fits}.
    The red cross marks the location in the CSM where the SN shock finally breaks out and the SN becomes observable. }
    \label{fig:CSM_density}
\end{figure}

The results of our modeling are shown in \autoref{fig:color_fits}.
The best-fit model is shown in the {middle} panel
of \autoref{fig:color_fits}, and the top panel shows the color coded
distribution of  $\ln\left(\chi^2/\chi_{\rm min}^2\right)$.
The best agreement with the observations is seen for 
$E_{\rm inj}=4.5\times10^{46}$ and $t_{\rm inj}=300\,{\rm days}$,
which would correspond to an eruption about $\sim200$ days before the SN\footnote{These parameters are very similar to the ones 
obtained from the same analysis for SN 2017eaw in \citet{morozova2020}. There, the best fit
energy is $E_{\rm inj}=5\times10^{46}\,{\rm erg}$, and the time
is $t_{\rm inj}=297\,{\rm d}$.}. 
While our search of archival data from ATLAS and ZTF shows no such eruption, it may be fainter than the detection limit due in part to the considerable extinction in the direction of SN\,2020fqv.
The error bars in the bottom panel
are constructed from all light curves that are within the white dashed
rectangle shown in the top panel. The dashed lines show the black body
magnitudes returned by \texttt{SNEC}, while the solid lines are computed from
\texttt{SYNOW} spectra. The difference between the two is generally minor,
apart from the late $g$-band light curve, where the iron line blanketing
effect is expected to be the strongest. Note that we did not perform
a broader study across different progenitor masses and final SN energies,
which could result in a slightly better fit.

\begin{figure}
  \centering
  \includegraphics[width=0.475\textwidth]{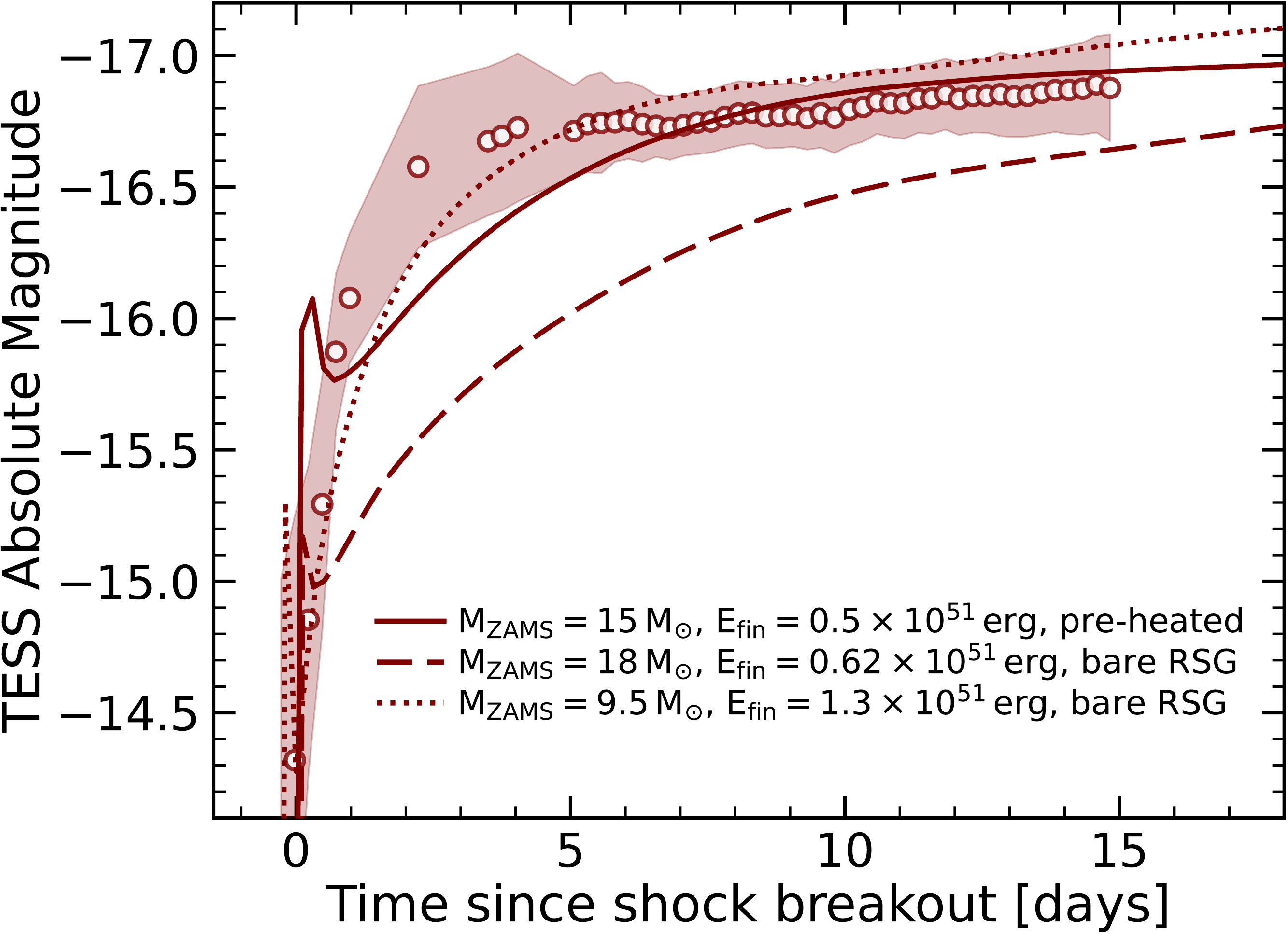}
  \caption{\textit{TESS} light curve of SN\,2020fqv compared to the
  numerical models. The solid line represents the best-fit model
  from \autoref{fig:color_fits}, in which the energy injection
  led to the formation of a CSM. The early spike seen in
  the model is a numerical artifact due to the fact that the photosphere is not
  resolved in our simulations during the first several hours of the light curve.
  Both dashed and dash-dotted
  lines represent the light curves of bare RSG models. The dashed
  line represents the model that provides the best fit for the 
  $r$-band magnitude during days $35-110$ of the light curve, while
  the dash-dotted line fits best the first $15$ days of the light curve.
  The corresponding light curves and data in $r$-band are shown
  in the inset.
  Even though the dash-dotted line fits well the early \textit{TESS} data,
  it cannot be regarded as a fit for the entire SN light curve.} 
  \label{fig:tess_fits}
\end{figure}

{The bottom panel of \autoref{fig:color_fits} justifies the need to 
pre-heat the RSG models by a weak shock wave before 
simulating the
SN explosion. In that panel we plot only $r$-band light 
curves, but the other filters would show a similar picture. The
red solid curve shows our best-fit model from the panel above, 
while the black curves show the `fits' that we obtained 
for the same data from
the bare RSG models that were not pre-heated and did not
form a dense CSM. Specifically, the dashed line shows the
model that fits best the $r$-band magnitude in days 
$35-110$ after the shock breakout, the dotted line fits the data
in days $0-20$ after the shock breakout, and the solid black
line aims to fit the entire data between day $0$ and day
$110$ of the light curve\footnote{All three black light
curves show the black body magnitudes returned by SNEC. 
However, as can be seen from the middle panel of
\autoref{fig:color_fits}, post-processing SNEC output
with SYNOW changes the $r$-band light curve only
slightly.}. The figure shows that none of the bare 
RSG models can fit the fast early rise of SN\,2020fqv light curve and its
plateau part at the same time. Pre-heating the model, even though
it does not provide the perfect fit to the data, changes the shape
of the simulated light curve that brings it in a better agreement with the data.}

In \autoref{fig:tess_fits}, we compare our models to the
\textit{TESS} data for SN\,2020fqv. The best-fit model from 
\autoref{fig:color_fits} is represented there by solid lines.
In that model, injection of $E_{\rm inj}=4.5\times10^{46}$
at the base of the hydrogen envelope led to the ejection
of the outer layers of the star and formation of the CSM.
For comparison, the dashed line represents the light curve
of the bare RSG model that can fit the $r$-band magnitude
of SN\,2020fqv during days $35-110$ of the light curve. 
The plots shows that pre-heating the model noticeably
improves its agreement with the data
in the early part of the light curve. At the same time, even the
pre-heated model does not rise sufficiently fast when
compared to the \textit{TESS} light curve, which suggests that further
modifications to the theoretical models should be considered.

On the other hand, it is possible to find a bare RSG model
that fits the early \textit{TESS} data without a 
CSM. 
The dotted line in \autoref{fig:tess_fits}
represents the model that fits best the first $20$ days 
of the $r$-band light curve, providing a good fit to the
TESS data as well (this model is the same as the one shown
by the dotted black line in the bottom panel of \autoref{fig:color_fits}).
This model has ZAMS mass of $9.5\,M_{\odot}$
and final energy of $1.3\times10^{51}\,{\rm erg}$.
However, as shown in \autoref{fig:color_fits}, this
model continues rising above the data and demonstrates too short
of a plateau to be regarded as a fitting model for SN\,2020fqv.
Indeed, for example, in \citet{pumo2017} and \citet{morozova2018} 
it was shown that the final energies needed
to reproduce plateau lengths of typical
SNe II-P lay below $\sim$$1.0\times10^{51}\,{\rm erg}$.
For this reason, we conclude that it is challenging to 
infer the SN parameters based on the early data only,
and it is important to take into account the entire light
curve.

From the best-fit models to the early-time light curve, we can infer the density profile of the progenitor star and the CSM, which is shown in \autoref{fig:CSM_density}.
The CSM is so optically thick that the shock breakout does not happen at the edge of the stellar envelope, but inside the CSM. 
The red cross in \autoref{fig:CSM_density} marks the location of the shock breakout, at about $9 \times 10^{13} \, \rm cm$, about $2.5 \times 10^{13} \, \rm cm$ (360 \rsun) above the stellar envelope. 
With the CSM, the shock breakout happens at a lower density than it would in a bare RSG envelope due to the more gradual density gradient in the CSM.
Thus, the shocked material is able to expand and cool faster, resulting in an early time excess flux compared to what is expected in a SN from a bare RSG progenitor. {This material is responsible for
filling the `gap' seen in the bottom panel of \autoref{fig:color_fits} between the dashed bare RSG model and the
data in the first $30$ days since shock breakout.}

Finally, it is informative to compare the characteristics of the model 
that fits SN\,2020fqv to a larger set of models studied in \citet{morozova2018}. 
In that work, the authors found numerical fits to the light curves of 20
well-studied SNe II-P by adding the CSM to their RSG models artificially in the
form of a dense wind. The CSM masses that were needed 
in order to fit the early data varied between
$0.003$ and $0.83\,M_{\odot}$, while the external radii of the
CSM varied between $700$ and $2200\,R_{\odot}$ for different
SNe.

\begin{figure}
    \centering
    \includegraphics[width=\linewidth]{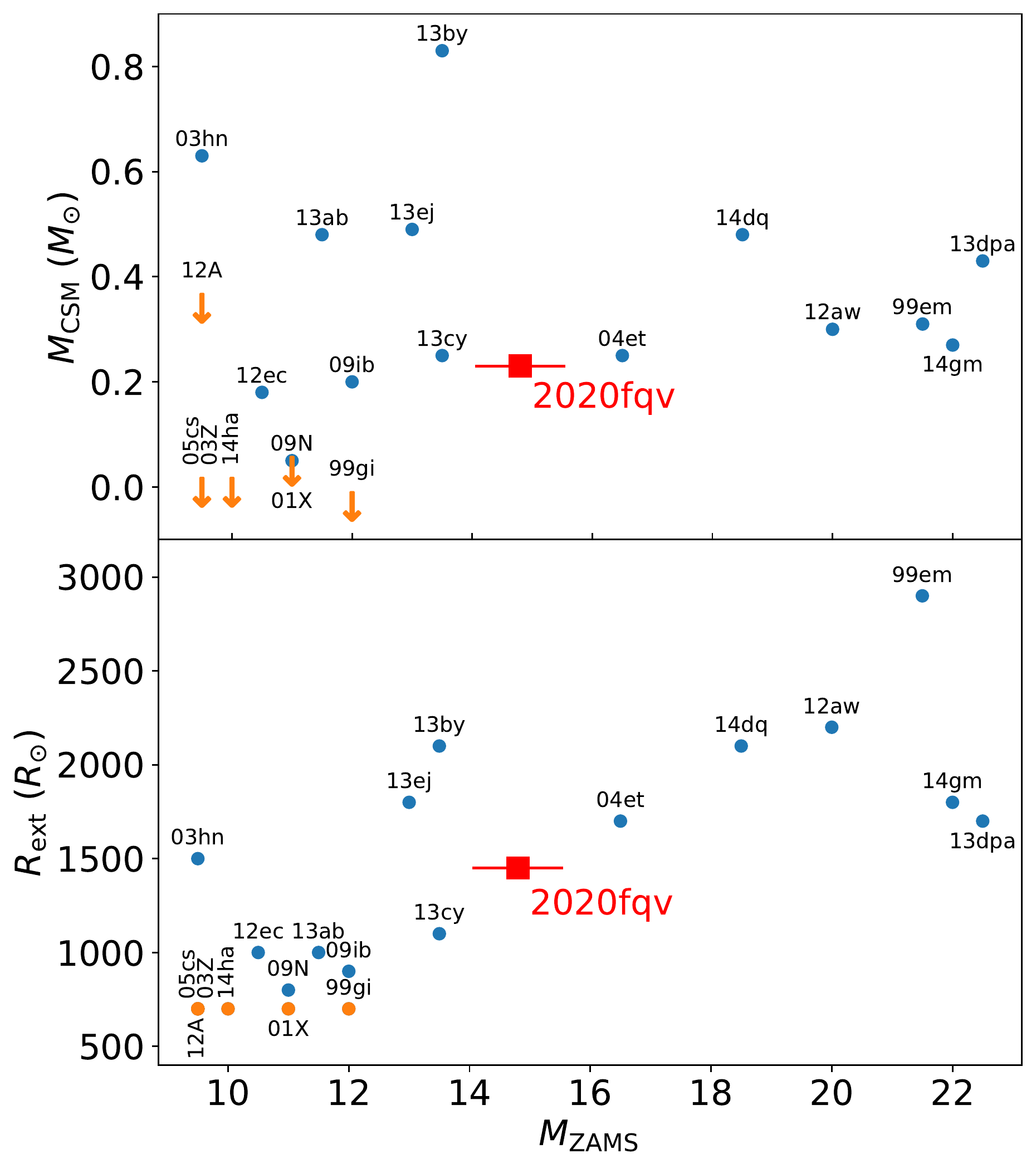}
    \caption{ Inferred CSM mass (\textbf{top}) and outer radius (\textbf{bottom}) as a function of $M_{\rm ZAMS}$ of SN\,2020fqv compared with other SNe~II-P from \protect\citet{morozova2018}. The error bar in $M_{\rm ZAMS}$ of SN\,2020fqv is from comparing different mass measurements in this work. Orange markers denote events with no detected CSM; $R_{\rm ext}$ measures the size of the progenitor for those events.}
    \label{fig:compare_IIP}
\end{figure}

We start from estimating the external radius of the CSM in
our best fit model for SN\,2020fqv. The radius of the original
RSG model is $841\,R_{\odot}$ (which is also the inner radius of the CSM), while the radius of our best
fit model prior to the 
SN explosion is $1450\,R_{\odot}$. This external radius
is in agreement with the CSM radii estimated in \citet{morozova2018}. Since the
best fit model for SN\,2020fqv corresponds to $t_{\rm inj}\approx 300$
days ($\approx200$ days between the pre-explosion outburst
and the SN), we estimate the velocity of CSM to be
$v_{\rm CSM}\approx 24\,{\rm km}\,{\rm s}^{-1}$.
The best fit model for SN\,2020fqv corresponds to pre-heating energy
$E_{\rm inj}=4.5\times10^{46}\,{\rm erg}$, but
only a fraction of this energy goes into the kinetic energy of the
outflow ($E_{\rm kin}\approx 1.4\times10^{45}\,{\rm erg}$ in our
simulation). Estimating the CSM mass from the formula
$2E_{\rm kin}/v_{\rm CSM}^2$ we obtain 
$M_{\rm CSM}\approx 0.23\,M_{\odot}$. This CSM mass
lays within the range of values obtained in \citet{morozova2018}, 
and it is very close
to the CSM mass obtained there for SN 2004et
($0.25\,M_{\odot}$).
\autoref{fig:compare_IIP} compares the CSM properties of SN\,2020fqv to those of SNe~II-P in \citet{morozova2018}; showing that the CSM around SN\,2020fqv is ordinary among the population of SNe~II-P. 

\subsubsection{Early-time spectroscopy}
\begin{figure}
    \centering
    \includegraphics[width=\linewidth]{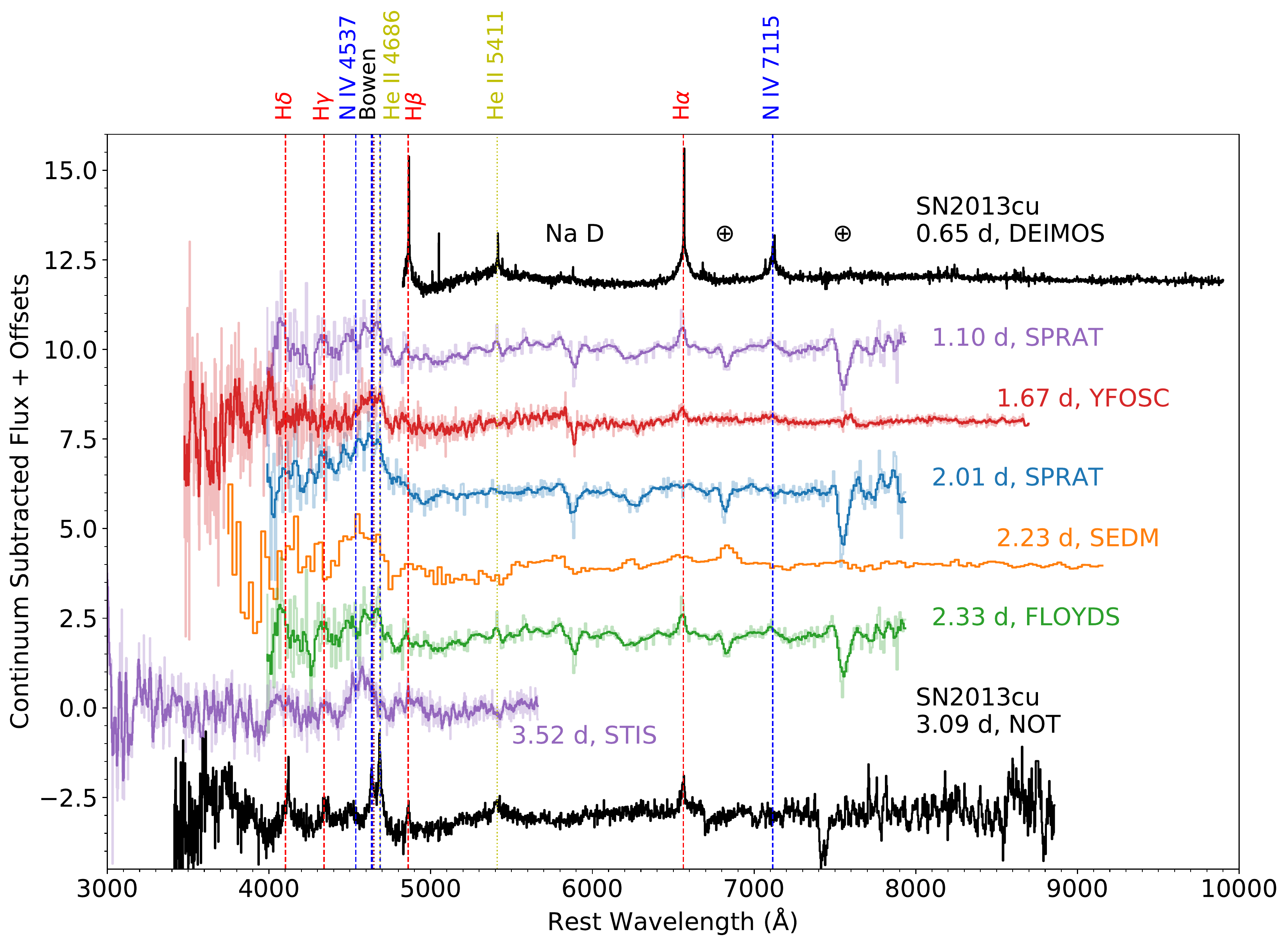}
    \caption{Early spectra of SN\,2020fqv from 1.1 to 3.52 d post-explosion, compared with the prototypical flash-ionization event SN\,2013cu \citep{galyam2014}.
    Solid lines are smoothed spectra; transparent lines are unsmoothed. 
    All spectra have been continuum subtracted to highlight emission lines.
    Line identifications are provided.
    The emission feature around 4600 \AA~ marked "Bowen" are the blend of \ion{C}{III} 4647, 4650~\AA\ and \ion{N}{III} 4634, 4641~\AA, excited likely via the Bowen fluorescence mechanism. }
    \label{fig:early_spec}
\end{figure}

\autoref{fig:early_spec} shows early time spectra from 1.1 to 3.52 d of SN\,2020fqv. 
For these spectra, a reddened black body continuum is fit and subtracted to highlight the emission lines.
We tentatively identify narrow H$\alpha$ emission; however, it is likely due to the host galaxy as it persisted throughout the plateau phase.  
No other Balmer lines are present.
The prominent emission feature that is likely from the SN is the emission around 4600 \AA. 
This feature appears in all the early-time spectra. 
\autoref{fig:uv_spec} clearly shows that it weakens quickly and completely disappears by 11 d. 
This emission feature is likely a blend between \ion{C}{III}, \ion{N}{III}, and \ion{He}{II}. 
The strength of these high-ionization lines relative to the Balmer series indicated that they are not excited thermally, but perhaps by the Bowen fluorescence mechanism \citep{bowen1935}.
In this process, the \ion{C}{III} and \ion{N}{III} ions are excited by the \ion{He}{II}~304~\AA\ emission, and cool via the \ion{C}{III}~4647,~4650~\AA\ and \ion{N}{III}~4634,~4641~\AA\ lines.  
However, we note that the STIS spectrum may require \ion{N}{IV} 4537 \AA\ emission to explain the emission feature. 
\autoref{fig:early_spec} also shows spectra at 0.65 and 3.09 d post-explosion of SN\,2013cu, the prototypical SN with narrow lines in early spectra, the so-called flash ionization features \citep{galyam2014}.
By comparison, it is clear that the narrow lines in SN\,2020fqv are much weaker that those in SN\,2013cu with no clear electron-scattering wings, pointing to weaker interactions. 
Because of the large optical depth of the CSM around SN\,2020fqv, these narrow lines must emerge from the outer part of the CSM, above the location of shock breakout shown in \autoref{fig:CSM_density} ($> 9\times 10^{13} \, \rm cm$).


\subsubsection{High-velocity component in the H$\alpha$ line}
\begin{figure}
    \centering
    \includegraphics[width = 0.7\linewidth]{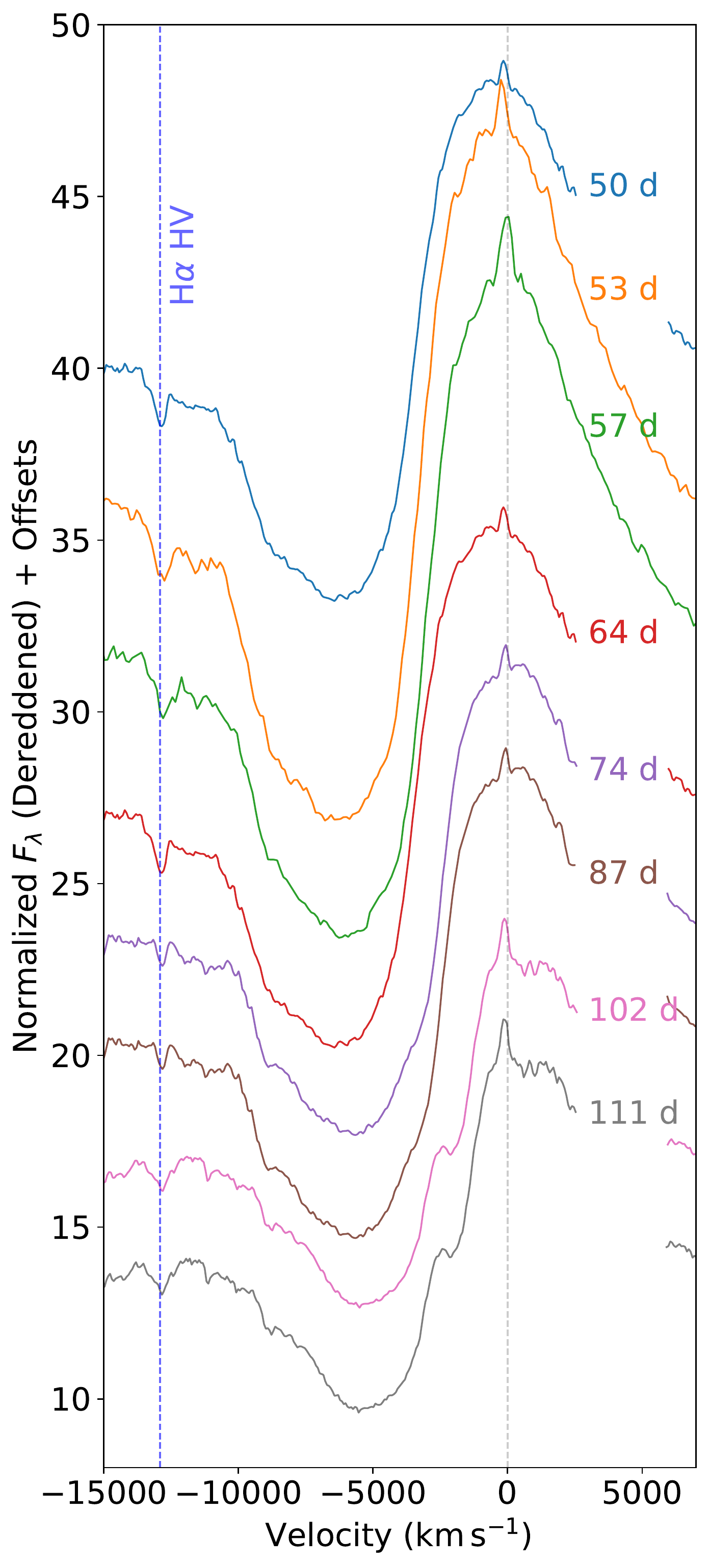}
    \caption{The profile of the H$\alpha$ line in the latter half of the plateau phase, plotted in velocity space. 
    The zero velocity is marked with a dashed grey line; the narrow component at this position is from the host galaxy. 
    We identify a persistent absorption at a constant velocity of $-1$2,900 $\rm km\,s^{-1}$ (marked with a dashed blue line), likely due to the outer ejecta being excited by ongoing CSM interactions. 
    This is in contrast with the velocity of the absorption trough of the P-Cygni profile, whose speed decreases towards the end of the plateau phase. }
    \label{fig:ha_hv}
\end{figure}

\autoref{fig:ha_hv} shows the evolution of the H$\alpha$ line profile from 50 to 111 d in the latter half of the plateau phase. 
The P-Cygni absorption trough slowed down from 6500 $\rm km\,s^{-1}$ to 5000 $\rm km\,s^{-1}$ as the photosphere receded into the slower ejecta.
We identified a persistent high-velocity (HV) absorption feature at about $-$12,900 $\rm km\,s^{-1}$.
This velocity corresponded to the wavelength of 6280 \AA, at which there was no other lines.
Such an absorption component could arise from an ongoing CSM interaction, as the high-energy photons from the interaction continuously excited the outer, fast-moving ejecta \citep{chugai2007}.
Because this feature persisted until the end of the plateau phase, it was not likely due to time-dependent effects \citep{dessart2008}.
Since this feature is persistent, it could not have come from the same inner CSM component responsible for the early flux excess and the early-time narrow emission lines.
It is a more extended CSM likely formed by the RSG wind. 

HV absorption components of strong spectral lines have been observed in a number of SNe II-P.
\cite{gutirrez2017} found that the absorption at around this wavelength is due to the \ion{Si}{ii}~6355~\AA\ line in the early plateau phase ($\lesssim$ 35 d), and due to the HV component of H$\alpha$ later in the plateau phase. 
However, the better place to clearly detect the HV component is the \ion{He}{i}~1.083~$\mu$m because there is less contamination from other lines at those wavelengths for SNe II-P. 
The HV component of the \ion{He}{i}~1.083~$\mu$m line has been detected in virtually all SNe II-P with near-IR spectroscopy (e.g. \citealp{tinyanont2019b, davis2019}), indicating that CSM interactions at this level is common.
In fact, \cite{davis2019} listed the presence of the HV component of the \ion{He}{i}~1.083~$\mu$m line as a feature distinguishing SNe II-P from II-L.

\subsection{Comparison with other SNe II with early UV spectra} \label{sec:UV_comp}
\begin{figure}
    \centering
    \includegraphics[width = \linewidth]{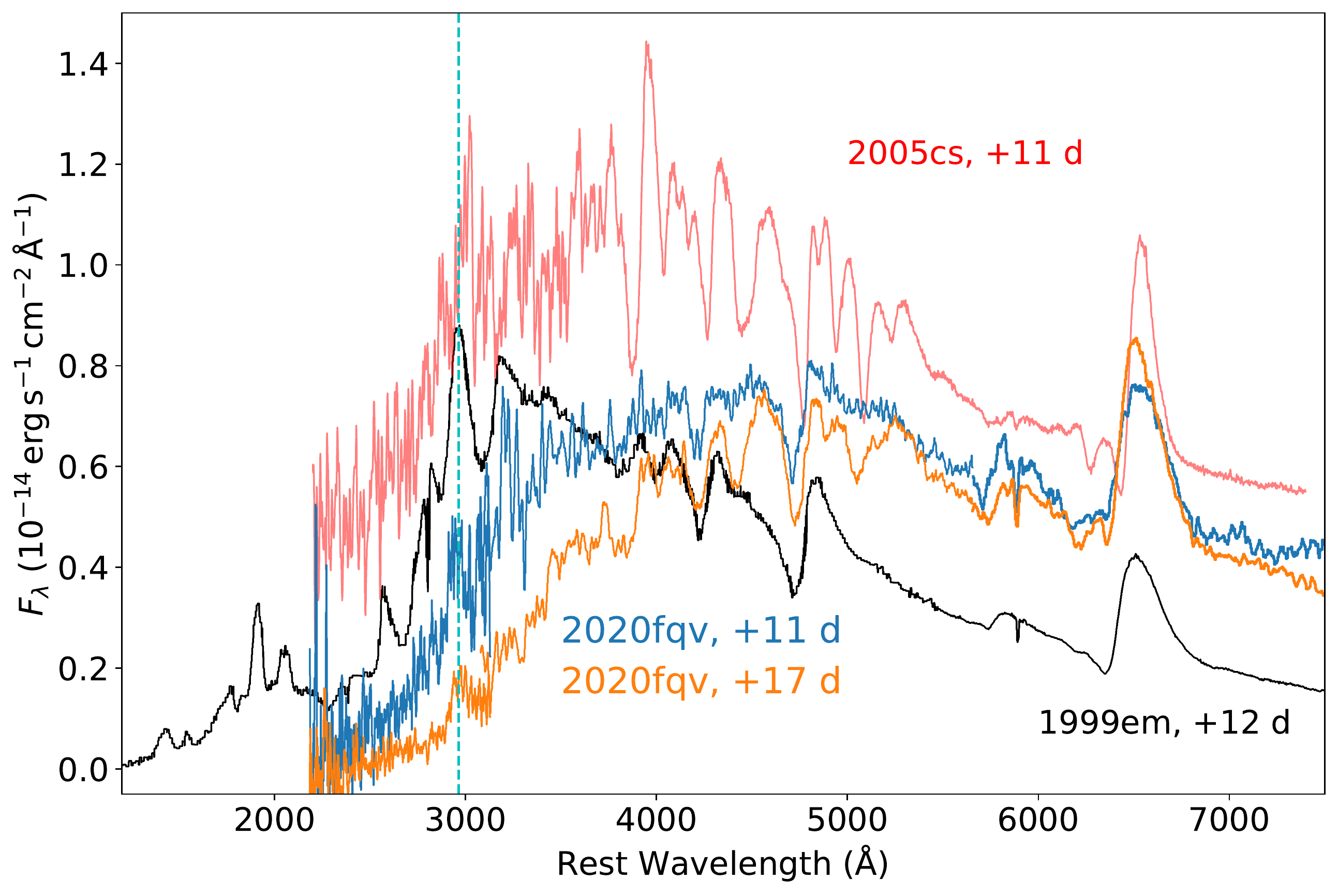}
    \caption{UV to optical spectra of SN\,2020fqv from 11 and 17 days post-explosion.
    Similar spectra of SNe\,1999em \citep{baron2000} and 2005cs \citep{bufano2009} are plotted for comparison. 
    The 2965 \AA\, feature due to Fe-line blanketing is marked by a dashed cyan line; the feature is visible in all spectra. 
    These are the only SNe II-P with high-quality early-time UV spectra available in the literature. 
    All spectra have been corrected for reddening. 
    The flux scale is based on the calibrated \textit{HST} spectra of SN\,2020fqv.
    The literature spectra are scaled for visualization.}
    \label{fig:uv_spec_comp}
\end{figure}

UV spectroscopic observations of SNe II remain sparse as only \textit{HST} STIS and COS have adequate sensitivities. 
The UV flux also fades quickly in SNe II-P due to Fe line blanketing.
\autoref{fig:uv_spec_comp} shows UV-to-optical spectra of SN\,2020fqv from 11 and 17 days post-explosion, in comparison to two other SNe II-P. 
SN\,1999em was observed with \textit{HST}/STIS at 12 days post-explosion \citep{baron2000}, while SN\,2005cs was observed with \textit{Swift} UV grism at 11 days post-explosion \citep{bufano2009}.
These are the only UV spectra of SNe II-P available in the literature at comparable epochs with comparable wavelength coverage. 
We note that SN\,2005ay also has UV spectroscopy at around 12 days post-explosion obtained by the \textit{Galaxy Evolution Explorer} (\textit{GALEX}) \citep{gal-yam2008}, but the spectra only cover up to 2900 \AA. 
Lastly, \cite{dhungana2016} presented several epochs of UV spectroscopy of SN\,2013ej, showing similar spectral shape and features as other SNe~II-P observed at similar epochs. 

The only discernible spectral feature of SN\,2020fqv in the UV is the feature at 2965 \AA, most visible in the +17 d spectrum. 
This peak is due to Fe line blanketing, absorbing the continuum flux on either side of it. 
The presence of this feature demonstrates that line blanketing starts to play a role already at +11 d for SN\,2020fqv.
This feature is also present in the spectrum of SN\,1999em; however, the UV flux in SN\,1999em is not yet as strongly absorbed as is the case in SN\,2020fqv. 
This may be because SN\,1999em synthesized less \nickel\ ($\sim$0.02 \msun; \citealp{elmhamdi2003}) compared to that of SN\,2020fqv (0.043 \msun). 
SN\,2005cs shows a similarly strong Fe line blanketing in the UV, even though it only synthesized 0.003--0.008 \msun\, of \nickel\ \citep{utrobin2008, pastorello2009}.
It is, however, a sub-luminous SN II-P resulting from a low mass ($\sim$9~\msun) progenitor star \citep{maund2005, li2006, smartt2009}. 
With the low ejecta mass, the spectral features emerge much earlier compared to SN\,2020fqv, including the Fe-line blanketing.
While early UV spectroscopy can directly measure the abundance of Fe-peak elements and the opacity evolution of the ejecta \citep[e.g.,][]{foley2013}, observations are presently sparse and a meaningful comparison cannot be made.  
Future \textit{HST}/STIS observations within $\sim$week of the explosion are needed to fully probe the landscape of early UV emissions in CCSNe.


\section{Conclusions}\label{sec:conclusion}

\begin{figure}
    \centering
    \includegraphics[width=\linewidth]{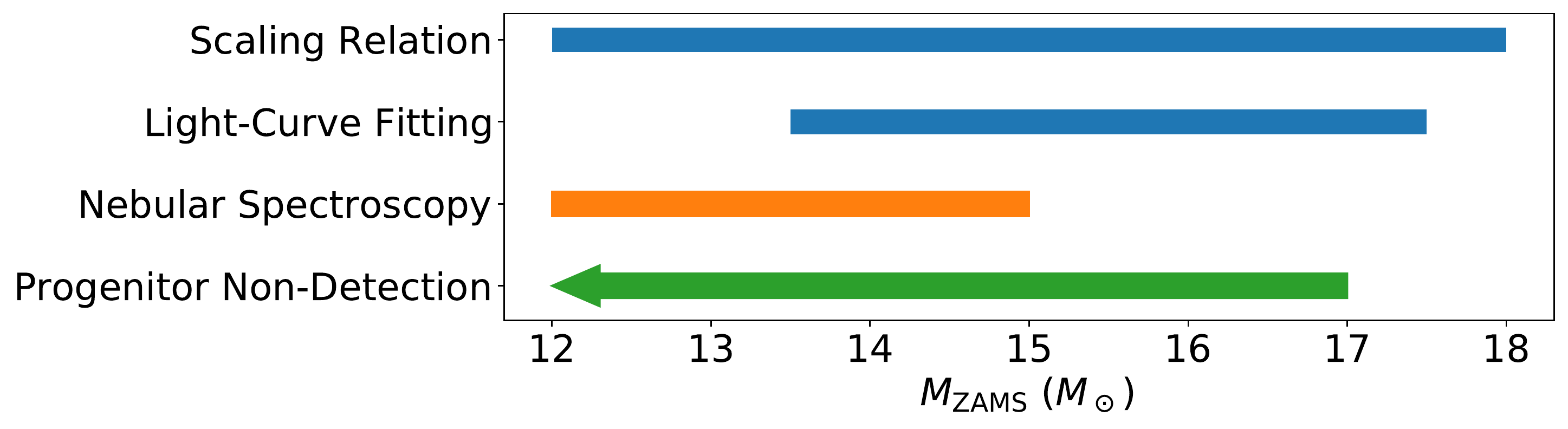}
    \caption{A summary of different progenitor mass estimates presented in this work. Different colors represent the different types of methods used to constrain the progenitor mass: blue is light-curve fitting, orange is nebular spectroscopy, and green is pre-explosion imaging. All measurements agree that SN\,2020fqv is an explosion of an average-mass RSG with $M_{\rm ZAMS}$ around 13.5--15 \msun.}
    \label{fig:prog_mass_comp}
\end{figure}

Despite being the most common subtype of CCSNe, Type II-P SNe still harbor many mysteries as recent observations uncover phenomena unexplained by simple models of RSG explosions in a vacuum. 
In this work, we present the explosion parameters of the nearby SN\,2020fqv.
With the high-cadence \textit{TESS} data, we are able to determine the explosion epoch to within 4 hours and monitor its early photometric evolution. 
From the bolometric light curves, we measure the mid-plateau (50 d post-explosion) luminosity of $L_{50} = (1.3 \pm 0.3) \times 10^{42} \, \rm erg\, s^{-1}$ and the plateau length of $t_p = 114 \pm 1 \, \rm d$.  
We determine the kinetic energy of the explosion to be $(4.1 \pm 0.1) \times 10^{50} \, \rm erg$ and the \nickel\ mass is $0.043 \pm 0.017$ \msun. 
The progenitor mass is obtained by four different methods: simple scaling relation; light-curve fitting; nebular spectroscopy; and pre-explosion imaging non-detection. 
\autoref{fig:prog_mass_comp} compares the ranges of progenitor mass estimates obtained from the four methods, pointing to a $\sim$13.5--15~\msun\, RSG as the progenitor of SN\,2020fqv. 
This is a typical mass for a RSG progenitor to SNe II-P \citep[e.g.,][]{smartt2009IIP}.

We then show that SN\,2020fqv exhibits many signatures of shock interactions with a CSM ejected prior to the explosion.
SN\,2020fqv rises to its peak luminosity too quickly to be explained by a typical RSG explosion model without CSM.
Light curve fitting shows that the rise can be explained by an interaction between the SN shock and the CSM ejected in a pre-SN eruption with $4.5 \times 10^{46} \, \rm erg$ injected into the base of the hydrogen envelope about 300 days pre-explosion.
The (unobserved) eruption ejects $\sim$0.23 \msun\ of materials in to the CSM with the maximum radius of $\sim$1450 \rsun. 
\autoref{fig:compare_IIP} shows that these CSM properties are typical among SNe~II-P.
Early spectra show narrow emission lines from high-ionization metal species (\ion{C}{III}, \ion{N}{III}, \ion{N}{IV}) and \ion{He}{II} around 4600~\AA\ from this CSM. 
Some of these lines may be excited by the Bowen fluorescence mechanism \citep{bowen1935}. 
Throughout the plateau phase, the H$\alpha$ line shows a persistent high-velocity component at a constant velocity of -12,900~$\rm km\, s^{-1}$, likely due to the continuous CSM interaction exciting the outermost (and fastest moving) layer of the ejecta. 

These observations add to the mounting evidence that RSGs can explode with a substantial amount of CSM, more than what is expected from a standard RSG wind ($\sim${}$10^{-6} \, M_{\odot} \rm \, yr^{-1}$ versus 0.23 \msun\ in less than a year observed here).
This indicates that the late-stage evolution of massive stars is more complicated than what is previously thought. 
Even for RSGs whose evolution is not significantly affected by binary interactions (though see \citealp{zapartas2019, zapartas2021}), mass loss due to the late-stage nuclear fusion can still produce a CSM with diverse properties that interacts with the SN shock at early times \citep{quataert2012,fuller2017, wu2021}. 
While studies of CSM properties with moderately-sized samples of SNe II-P are present \citep[e.g.,][]{morozova2018}, the Vera Rubin Observatory will be able to produce multi-band light curves of all SNe II-P out to about 400 Mpc, allowing for a truly systematic study of CSM interactions around SNe II-P. 
With the pre-detection of SNe II-P progenitors reliant on serendipitous pre-explosion \textit{HST} imaging, if a connection between the CSM properties and progenitor properties can be made, this would greatly enhance our capability to study the late-stage evolution of RSGs.

\section*{Acknowledgements}

We thank Thomas de Jaeger for a discussion on the color of Type II SNe
and for providing us some data.

The script to colorize the single-band pre-explosion \textit{HST}
image is based on a script written by M. Durbin, which can be found
here:
\url{https://gist.github.com/meredith-durbin/c05bba9490017ad8e5bb7fd2d9774d04}.
This work has made use of the SVO Filter Profile Service
(http://svo2.cab.inta-csic.es/theory/fps/) supported from the Spanish
MINECO through grant AYA2017-84089.

The UCSC team is supported in part by NASA grant 80NSSC20K0953, NSF
grant AST-1815935, the Gordon \& Betty Moore Foundation, the
Heising-Simons Foundation, and by a fellowship from the David and
Lucile Packard Foundation to R.J.F.  M.R.S.\ is supported by the NSF
Graduate Research Fellowship Program Under grant 1842400. D.A.C.\
acknowledges support from the National Science Foundation Graduate
Research Fellowship under Grant DGE1339067. Support for this work was
provided by NASA through the NASA Hubble Fellowship grant
HF2-51462.001 awarded by the Space Telescope Science Institute, which
is operated by the Association of Universities for Research in
Astronomy, Inc., for NASA, under contract NAS5-26555.

This work is based on observations made with the NASA/ESA {\it Hubble
Space Telescope} under program number GO-15876 and data from program
number GO-5446 obtained from the data archive at the Space Telescope
Science Institute.  Support for program GO-15876 was provided by NASA
through a grant from STScI, which is operated by AURA, Inc., under
NASA contract NAS 5-26555.  Additional support was provided through
NASA grants in support of {\it Hubble Space Telescope} programs
GO-15889 and GO-16075.

A.G.\ is supported by the National Science Foundation Graduate
Research Fellowship Program under Grant No.\ DGE-1746047. A.G.\ also
acknowledges funding from the Center for Astrophysical Surveys
Fellowship at UIUC/NCSA and the Illinois Distinguished Fellowship.
C.D.K.\ acknowledges support through NASA grants in support of {\it Hubble Space Telescope} programme AR-16136.
W.J.-G.\ is supported by the National Science Foundation Graduate
Research Fellowship Program under Grant No.\ DGE-1842165 and the IDEAS
Fellowship Program at Northwestern University.
I.P.-F.\ acknowledges support from the Spanish State Research Agency
(AEI) under grant numbers ESP2017-86852-C4-2-R and
PID2019-105552RB-C43.
F.P.\ acknowledges support from the Spanish State Research Agency
(AEI) under grant number PID2019-105552RB-C43.
Q.W.\ acknowledges financial support provided by the STScI Director's
Discretionary Fund.

This paper includes data collected by the TESS mission. Funding for
the TESS mission is provided by the NASA's Science Mission
Directorate.

Some of the data presented herein were obtained at the W.\ M.\ Keck
Observatory, which is operated as a scientific partnership among the
California Institute of Technology, the University of California, and
NASA. The Observatory was made possible by the generous financial
support of the W.\ M.\ Keck Foundation.

This work is based in part on observations obtained at the international Gemini
Observatory, a program of NSF's NOIRLab (programs GN-2020A-Q-134 and
GS-2020A-Q-128), which is managed by the Association of Universities
for Research in Astronomy (AURA) under a cooperative agreement with
the National Science Foundation. on behalf of the Gemini Observatory
partnership: the National Science Foundation (United States), National
Research Council (Canada), Agencia Nacional de Investigaci\'{o}n y
Desarrollo (Chile), Ministerio de Ciencia, Tecnolog\'{i}a e
Innovaci\'{o}n (Argentina), Minist\'{e}rio da Ci\^{e}ncia, Tecnologia,
Inova\c{c}\~{o}es e Comunica\c{c}\~{o}es (Brazil), and Korea Astronomy
and Space Science Institute (Republic of Korea).

This work was enabled by observations made from the Gemini North and
Keck telescopes, located within the Maunakea Science Reserve and
adjacent to the summit of Maunakea. The authors wish to recognize and
acknowledge the very significant cultural role and reverence that the
summit of Maunakea has always had within the indigenous Hawaiian
community. We are grateful for the privilege of observing the Universe
from a place that is unique in both its astronomical quality and its
cultural significance.

A major upgrade of the Kast spectrograph on the Shane 3~m telescope at
Lick Observatory was made possible through generous gifts from the
Heising-Simons Foundation as well as William and Marina Kast. Research
at Lick Observatory is partially supported by a generous gift from
Google.

This work makes use of observations from the Las Cumbres Observatory
global telescope network following the approved NOIRLab programs
2020A-0196, 2020A-0334, 2020B-0250, 2020B-0256, 2021A-0135, and
2021A-0239.  LCO telescope time was granted by NOIRLab through the
Mid-Scale Innovations Program (MSIP). MSIP is funded by NSF.

The Liverpool Telescope is operated on the island of La Palma by
Liverpool John Moores University in the Spanish Observatorio del Roque
de los Muchachos of the Instituto de Astrofisica de Canarias with
financial support from the UK Science and Technology Facilities
Council.

This work is based in part on observations obtained with the Samuel Oschin 48-inch
Telescope at the Palomar Observatory as part of the Zwicky Transient
Facility project. ZTF is supported by the NSF under grant AST-1440341
and a collaboration including Caltech, IPAC, the Weizmann Institute
for Science, the Oskar Klein Center at Stockholm University, the
University of Maryland, the University of Washington, Deutsches
Elektronen-Synchrotron and Humboldt University, Los Alamos National
Laboratories, the TANGO Consortium of Taiwan, the University of
Wisconsin at Milwaukee, and the Lawrence Berkeley National
Laboratory. Operations are conducted by the Caltech Optical
Observatories (COO), the Infrared Processing and Analysis Center
(IPAC), and the University of Washington (UW).

This work has made use of data from the Asteroid Terrestrial-impact Last Alert System (ATLAS) project. The Asteroid Terrestrial-impact Last Alert System (ATLAS) project is primarily funded to search for near earth asteroids through NASA grants NN12AR55G, 80NSSC18K0284, and 80NSSC18K1575; byproducts of the NEO search include images and catalogs from the survey area. This work was partially funded by Kepler/K2 grant J1944/80NSSC19K0112 and HST GO-15889, and STFC grants ST/T000198/1 and ST/S006109/1. The ATLAS science products have been made possible through the contributions of the University of Hawaii Institute for Astronomy, the Queen~s University Belfast, the Space Telescope Science Institute, the South African Astronomical Observatory, and The Millennium Institute of Astrophysics (MAS), Chile.

The Pan-STARRS1 Surveys (PS1) and the PS1 public science archive have
been made possible through contributions by the Institute for
Astronomy, the University of Hawaii, the Pan-STARRS Project Office,
the Max-Planck Society and its participating institutes, the Max
Planck Institute for Astronomy, Heidelberg and the Max Planck
Institute for Extraterrestrial Physics, Garching, The Johns Hopkins
University, Durham University, the University of Edinburgh, the
Queen's University Belfast, the Harvard-Smithsonian Center for
Astrophysics, the Las Cumbres Observatory Global Telescope Network
Incorporated, the National Central University of Taiwan, STScI, NASA
under grant NNX08AR22G issued through the Planetary Science Division
of the NASA Science Mission Directorate, NSF grant AST-1238877, the
University of Maryland, Eotvos Lorand University (ELTE), the Los
Alamos National Laboratory, and the Gordon and Betty Moore
Foundation. Pan-STARRS is a project of the Institute for Astronomy of
the University of Hawaii, and is supported by the NASA SSO Near Earth
Observation Program under grants 80NSSC18K0971, NNX14AM74G,
NNX12AR65G, NNX13AQ47G, NNX08AR22G, 20-YORPD 20\_2-0014 and by the
State of Hawaii.

Palomar Gattini-IR (PGIR) is generously funded by Caltech, Australian National University, the Mt Cuba Foundation, the Heising Simons Foundation, and the Binational Science Foundation. PGIR is a collaborative project among Caltech, Australian National University, University of New South Wales, Columbia University, and the Weizmann Institute of Science. MMK acknowledges generous support from the David and Lucille Packard Foundation. MMK and Eran Ofek acknowledge the US-Israel Bi-national Science Foundation Grant 2016227. MMK and Jeno L Sokoloski acknowledge the Heising-Simons foundation for support via a Scialog fellowship of the Research Corporation. MMK and AMM acknowledge the Mt Cuba foundation.

\section*{Data Availability}

All spectra presented in this work are available via WISEReP. 
Other data and data analysis scripts can be obtained from the corresponding author upon a reasonable request.



\bibliographystyle{mnras}
\bibliography{SN2020fqv} 




\appendix


\section{Detailed Data Analysis Figures}

\begin{figure}
    \centering
    \includegraphics[width=\linewidth]{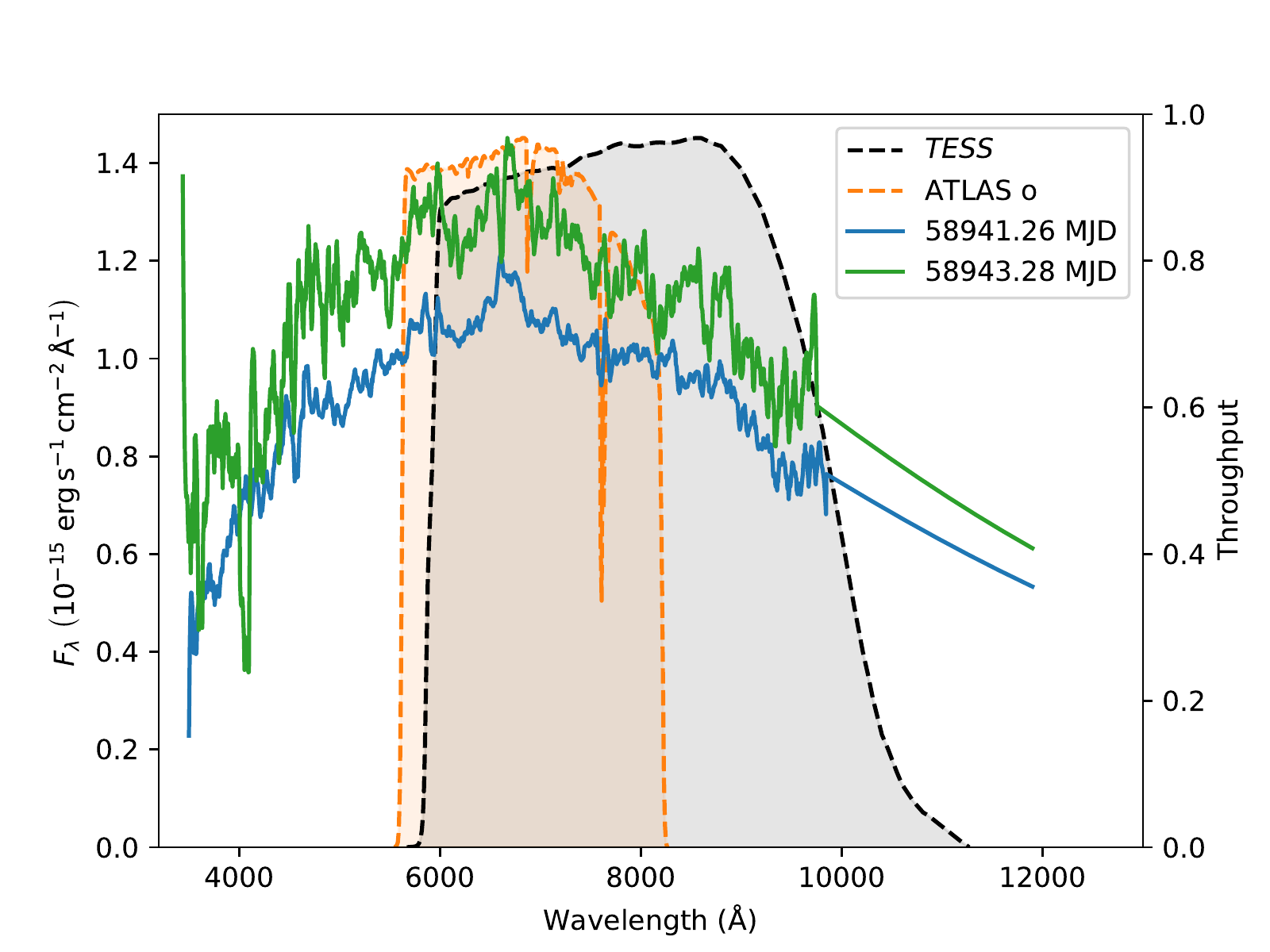}
    \caption{Flux calibrated FLOYDS spectra used to calculate the \textit{TESS} synthetic photometry. We extend the spectra to 1200~\AA\ by appending the best fitting black body models, to fully cover the TESS bandpass. The \textit{TESS} and ATLAS $o$ band passes that are available from SVO are overlaid. 
    }
    \label{fig:tess_flux_cal}
\end{figure}


\begin{figure}
    \centering
    \includegraphics[width=\linewidth]{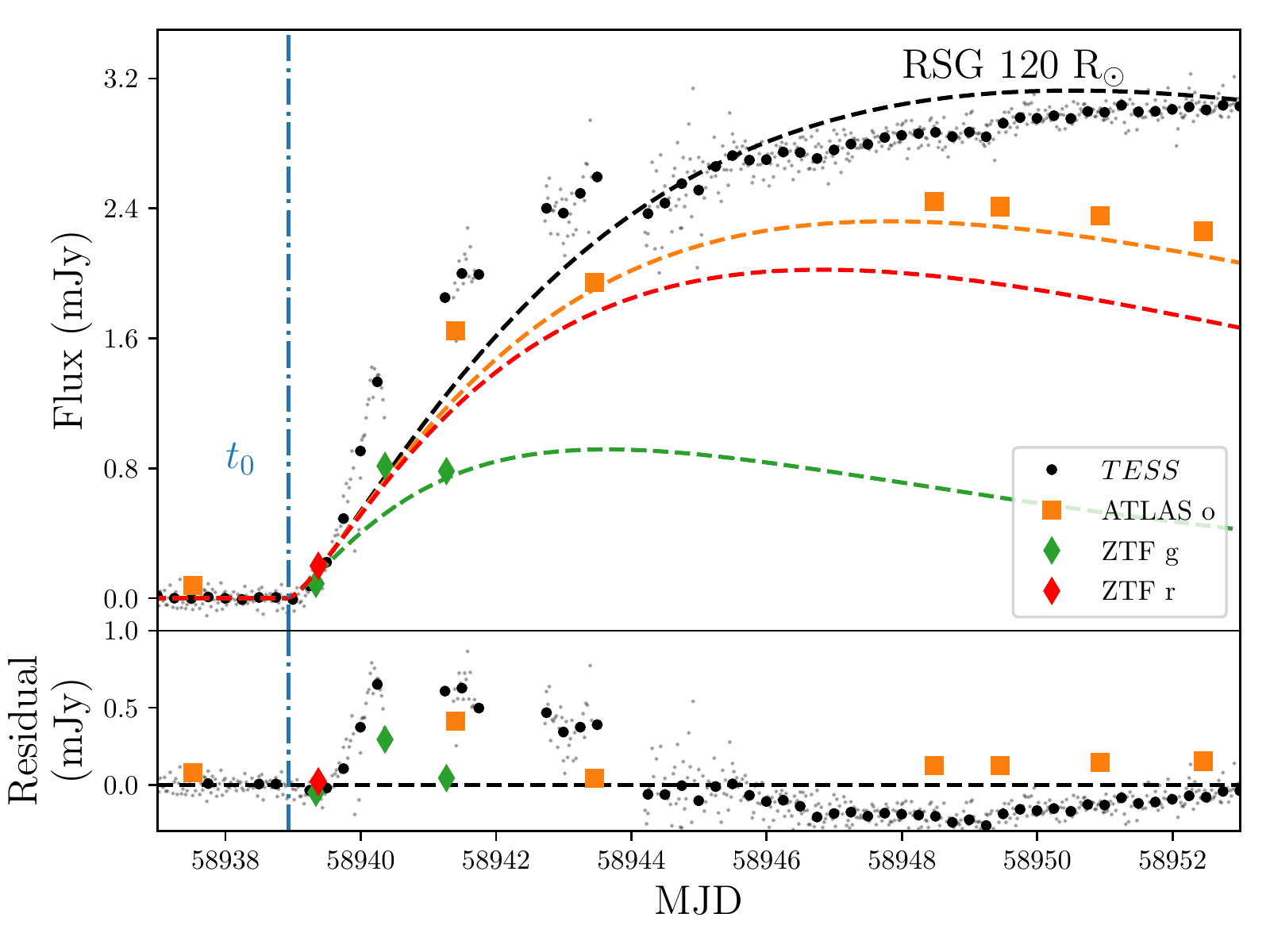}
    \includegraphics[width=\linewidth]{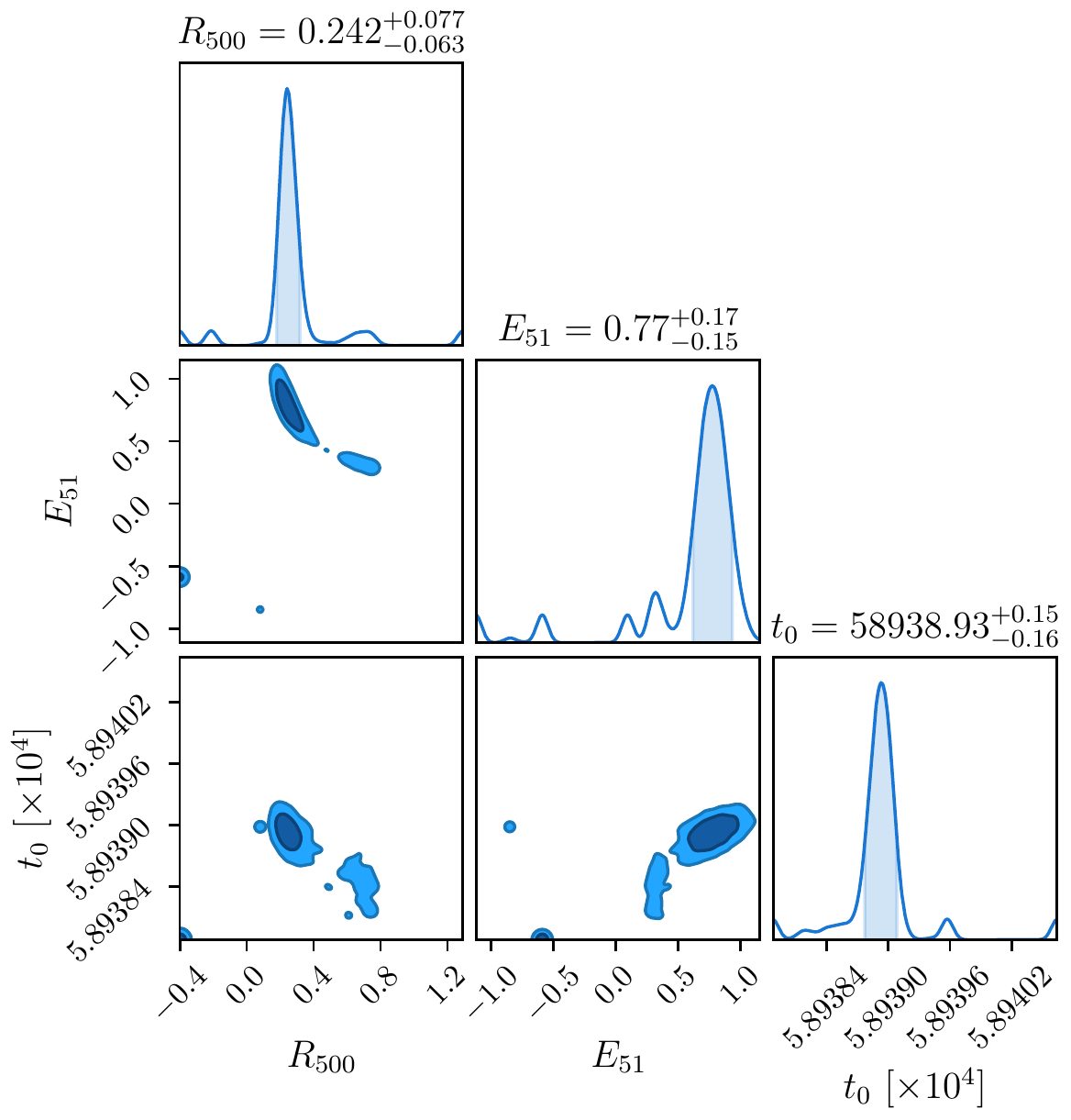}
    \caption{\citet{Nakar2010} analytical RSG model with a progenitor mass of 
    $15\rm ~M_\odot$ fit to early photometry. \textbf{Upper:} Early 
    \textit{TESS}, ATLAS, and ZTF light curves with corresponding light curves derived from the best fit parameters. \textbf{Lower:} Parameter distributions from the simultaneous \texttt{emcee} fit to all early photometry, made with \texttt{chainconsumer} \citep{Hinton2016}. Although the parameters are well constrained the enhanced brightness at early times due to CSM interaction produces unrealistic physical parameters, therefore, we only use the explosion time, $t_0$, from this fit.}
    \label{fig:nakar_fit}
\end{figure}

\begin{figure}
    \centering
    \includegraphics[width=\linewidth]{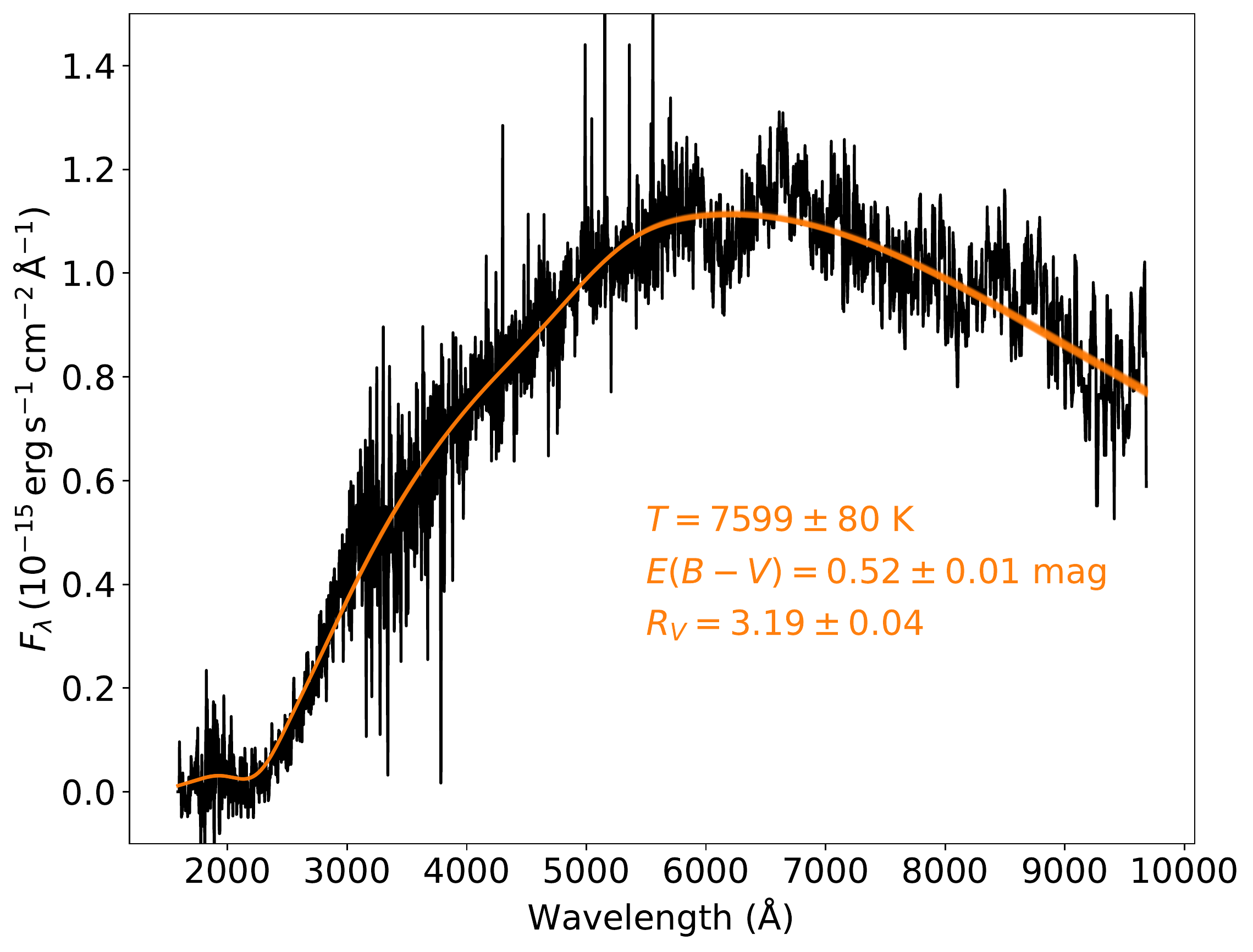}
    \caption{\textit{HST} and ground-based spectroscopy from 4 days post-explosion, where the SN emission is assumed to be well-described by a black body. We fit this spectrum with a black body model to constrain the dust extinction parameters. The resulting best-fit values are $E(B-V) = 0.52 \pm 0.01$ mag and $R_V = 3.19 \pm 0.04$.}
    \label{fig:dust_fit}
\end{figure}



\bsp	
\label{lastpage}
\end{document}